\documentclass[1p,times]{elsarticle} % compile w.o. figures
\usepackage{hyperref} % cross-reference
\usepackage[export]{adjustbox} % right-align minipage figures
\usepackage[labelfont=bf,justification=raggedright,singlelinecheck=false]{caption} % figure/table caption 
\usepackage[T1]{fontenc} % special characters(1/2)
\usepackage[utf8]{inputenc} % bibtex special characters(2/2)
\DeclareUnicodeCharacter{03BC}{\ensuremath{\mu}} % remove error! Package inputenc Error: Unicode char μ (U+3BC)
\DeclareUnicodeCharacter{00A0}{~} % remove error! Package inputenc Error: Unicode char \u8: not set up for use with LaTeX
\usepackage{graphicx}
\usepackage{amssymb}
\usepackage{lineno}

\usepackage{upgreek}
\usepackage{amsmath}
\usepackage{color}

\newcommand{\REV}[1]{{\color{black} #1}}

\journal{Biophysical Journal}

\begin{document}

\begin{frontmatter}



\title{Spatio-temporal dynamics of dilute red blood cell suspensions in a microchannel flow at low Reynolds number}

\author[1]{Qi Zhou\fnref{fn1}}
\author[2]{Joana Fidalgo\fnref{fn1}}
\author[1]{Lavinia Calvi}
\author[3]{Miguel O. Bernabeu}
\author[4]{Peter R. Hoskins}
\author[2]{M\'onica S. N. Oliveira\corref{cor1}}
\author[1]{Timm Kr\"uger\corref{cor1}}
\address[1]{School of Engineering, Institute for Multiscale Thermofluids, University of Edinburgh, Edinburgh EH9
3FB, UK}
\address[2]{James Weir Fluids Laboratory, Department of Mechanical and Aerospace Engineering, University of
Strathclyde, Glasgow G1 1XJ, UK}
\address[3]{Centre for Medical Informatics, Usher Institute, University of Edinburgh, Edinburgh EH16 4UX, UK}
\address[4]{Centre for Cardiovascular Science, University of Edinburgh, Edinburgh EH16 4SB, UK}
\fntext[fn1]{These authors contribute equally to this work.}
\cortext[cor1]{Corresponding author e-mail: monica.oliveira@strath.ac.uk or timm.krueger@ed.ac.uk}

\begin{abstract}
Microfluidic technologies are commonly used for the manipulation of red blood cell (RBC) suspensions and analyses of flow-mediated biomechanics. 
To enhance the performance of microfluidic devices, understanding the dynamics of the suspensions processed within is crucial.
We report novel aspects of the spatio-temporal dynamics of RBC suspensions flowing through a typical microchannel at low Reynolds number.
Through experiments with dilute RBC suspensions, we find an off-centre two-peak (OCTP) profile of cells contrary to the centralised distribution commonly reported for low-inertia flows. This is reminiscent of the well-known ``tubular pinch effect'' which arises from inertial effects. However, given the conditions of negligible inertia in our experiments, an alternative explanation is needed for this OCTP profile.
Our massively-parallel simulations of RBC flow in real-size microfluidic dimensions using the immersed-boundary-lattice-Boltzmann method (IB-LBM) confirm the experimental findings and elucidate the underlying mechanism for the counterintuitive RBC pattern.
By analysing the RBC migration and cell-free layer (CFL) development within a high-aspect-ratio channel, we show that such a distribution is co-determined by the spatial decay of hydrodynamic lift and the global deficiency of cell dispersion in dilute suspensions.
We find a CFL development length greater than 46 and 28 hydraulic diameters in the experiment and simulation, respectively, exceeding typical lengths of microfluidic designs.
Our work highlights the key role of transient cell distribution in dilute suspensions, which may negatively affect the reliability of experimental results if not taken into account.
\end{abstract}

\end{frontmatter}

\section{Introduction}
\label{Introduction}

Red blood cells (RBCs) occupy around 45\% of the blood volume and play a crucial role in maintaining the normal function of tissues and organs in humans and animals.
They transport oxygen and carbon dioxide, regulate blood viscosity, and affect the circulation of nutrients and immune cells.
Pioneered by Chien and Skalak in the 1960s--1980s \cite{chien_blood_1967, chien_viscoelastic_1975, skalak_rheology_1982}, there has been extensive research carried out theoretically, experimentally and numerically to elucidate  the behaviour of single RBC and characterise the dynamics of RBC suspensions (see recent reviews \cite{vlahovska_flow_2013, abreu_fluid_2014, freund_numerical_2014, geislinger_hydrodynamic_2014, yazdani_dynamic_2016}).
However, despite major progress, a rigorous and quantitative connection between microscale RBC dynamics and macroscale haemorheology is still lacking \cite{gompper_modeling_2015}.

Up to date, various factors affecting blood microrheology have been determined by theoretical/numerical models of RBCs or biomimetic vesicles, including (but not limited to): (i) viscosity contrast (between inner and outer fluids) \cite{danker_rheology_2007}, deformability \cite{kruger_crossover_2013}, orientation \cite{cordasco_orbital_2013} and initial position \cite{guckenberger_numerical-experimental_2018} of the cell; (ii) shear component (distinct in linear/quadratic flows) \cite{olla_behavior_2000}, suspending viscosity \cite{zhang_effect_2011}, flowline curvature \cite{ghigliotti_vesicle_2011} and wall confinement \cite{kaoui_two-dimensional_2011} of the flow; (iii) two-body or multi-body hydrodynamic interactions between cells \cite{aouane_hydrodynamic_2017}.
Most studies focus on the individual dynamics/pairwise interaction of cells or the characteristic signatures of the overall suspension system, e.g., effective viscosity and normal stress differences.

Much less attention has been paid to the collective behaviour of cells, e.g., their spatio-temporal organisation or local microstructures, at a realistic length scale in three-dimensional space.
Common reasons for avoiding such studies include their complexity of analytical modelling or prohibitiveness of computational cost.
In fact, the RBC behaviour at sub-suspension level plays a key role in bridging the RBC dynamics and haemorheology. From an experimental perspective, the maturing of microfluidic technologies over the last two decades has provided a powerful tool to explore the otherwise inaccessible behaviour of complex biofluids at the microscale.
It is now widely employed for blood flow experiments aimed at various applications \cite{toner_blood---chip_2005, tomaiuolo_biomechanical_2014, sebastian_microfluidics_2018}.
For practical reasons in microfluidics, dilute (\REV{concentration $\phi < 10$\%}) or semi-dilute RBC suspensions are commonly used, where the clogging of samples is minimal and the measurements of individual cells are quantitatively accurate \cite{lima_radial_2008, yaginuma_human_2013, roman_going_2016}.
Through observation of dilute/semi-dilute RBCs in polydimethylsiloxane (PDMS) microfluidic channels, many intriguing properties of RBC flow at the microscale have been demonstrated, such as uneven cell-plasma partitioning \cite{yang_microfluidic_2006}, long-range hydrodynamic interaction \cite{kantsler_dynamics_2008}, wall-induced lift of cells in tumbling motion \cite{grandchamp_lift_2013}, cytoplasmic viscosity-promoted polylobes on cell membrane \cite{lanotte_red_2016} and diffusion-governed collective spreading of cell suspensions \cite{chuang_collective_2018}.

From experiments with a semi-dilute RBC suspension ($\phi = 16$\%), Sherwood et al. \cite{sherwood_spatial_2014} revealed strongly heterogeneous distribution of cells, which presented locally skewed rather than axisymmetric haematocrit ($Ht$) profiles.
With much lower RBC concentration ($\phi \leq 5$\%), Shen et al. \cite{shen_inversion_2016} found an even more heterogeneous RBC distribution eventually leading to inversion of the classic Zweifach-Fung effect (governing cell partitioning) downstream of bifurcations.
\REV{More recently, Iss et al. \cite{iss_self-organization_2019} and Abay et al. \cite{abay_cross-sectional_2020} discovered emergent self-organisation or focusing behaviour of cells in dilute RBC suspensions arising from geometric features of the microfluidic channel, such as confinement and constriction.} 
These studies imply that the spatial arrangement of RBCs in microchannels can be highly dynamical and varies from one location to another.
Such local characteristics, if ignored, may give rise to inaccurate estimation of the haemodynamics in the microfluidic channels, and affect the physiological relevance of the in vitro experiments designed to imitate in vivo systems.

The present work aims to elucidate the spatio-temporal distribution of RBCs in a dilute suspension flowing through a typical microfluidic channel of rectangular cross-section.
For a dilute suspension ($Ht \leq$ 1\%) of horse RBCs in the microchannel, cell density profiles are obtained by a direct cell-counting routine developed \textit{in-house}. Resultantly, we observe a counterintuitive cell distribution with off-centre two-peak ordering rather than a single peak near the centreline, even in the negligible inertia regime.
Immersed-boundary-lattice-Boltzmann simulations performed for an equivalent system confirm the experimental finding and provide insight into the underlying mechanisms. The paper is organised as follows: the experimental and numerical methods are introduced in Sec.~\emph{\nameref{Materials and Methods}}.
In Sec.~\emph{\nameref{Counterintuitive RBC distribution}}, cell density profiles from experiments and simulations are presented.
Sec.~\emph{\nameref{Flow properties and RBC dynamics}} provides analysis of the observed cross-streamline migration and explores the mechanism behind the density profiles.
Sec.~\emph{\nameref{Discussion}} compares the present study with existing works and states the key implications of our findings.
Sec.~\emph{\nameref{Conclusion}} concludes the paper.

\section{Materials and Methods}
\label{Materials and Methods}

The materials and methods used for the experiments and the simulations involved in this study are outlined in Sec.~\emph{\nameref{Microfluidic experiments}} and Sec.~\emph{\nameref{Computational modelling}}, respectively.
More details are provided in the Supplemental Information (SI).

\subsection{Microfluidic experiments} 
\label{Microfluidic experiments}

RBC suspensions in Dextran40 were prepared using horse blood (see Fig.~\ref{fig:FigS1}, SI for Photomicrography of the RBC suspension) provided by TCS Biosciences, in anticoagulant EDTA (1.5 mg/mL). The procedures for sample processing are described in Fig.~\ref{fig:FigS2} and Sec.~\emph{\nameref{Preparation of RBC suspensions, SI}}. The steady shear rheology of Dextran 40 was determined by a rotational rheometer (DHR2, TA Instruments, with minimum resolvable torque 0.1 nNm) using a cone-plate geometry of 60 mm diameter and angle $\theta$ = 1$^\circ$ (CP60/1), 30 $\upmu$m truncation gap. The density and shear viscosity values used to calculate dimensionless numbers (e.g., Reynolds number $Re$ and Capillary number $Ca$) for experiments are 1.07 g/mL and 5.0 mPa~s, respectively.

\begin{figure*}[!t]
\centering
\includegraphics[width=1.0\linewidth]{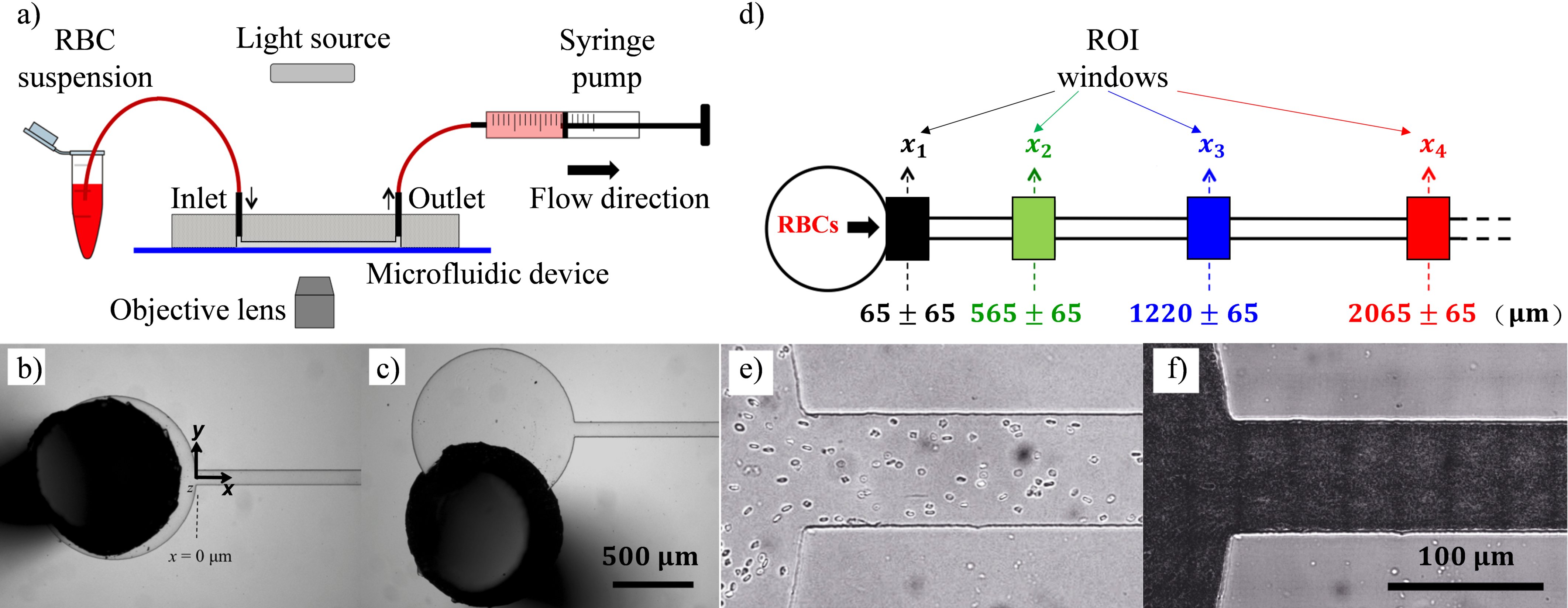}
\caption{\label{fig:Fig1} (a) Schematic of the experimental setup. (b--c) Inlet configuration of the microfluidic device, showing optical images of the flow entry region where the inlet port is (b) aligned and (c) misaligned relative to the channel centreline. In (b), the coordinate axes are shown. (b--c) share the same scale bar of 500 $\upmu$m. (d) Schematic showing designated regions of interest (ROIs) for the measurement of probability density distribution (PDD) of RBCs in the suspension. (e) An individual frame of RBCs flowing in the microchannel and (f) a composite image combining 300 frames of the same region, obtained using the Z-projection method based on minimum intensity. In (e--f), $Q = 0.2\,\upmu$L/min, and they share the same scale bar of 100 $\upmu$m.}
\end{figure*}

The microfluidic device was manufactured by soft-lithography using an SU8 mould produced by photo-lithography (microLIQUID). The microfluidic channel was made in PDMS (polydimethylsiloxane, Sylgard 184, Dow Corning) and bonded to a glass slide substrate with oxygen plasma (Zepto, Diener). The channel has a rectangular cross-section, with the width and depth dimensions ($W=87\,\upmu$m, $H=31\,\upmu$m) determined by optical microscopy. Fig.~\ref{fig:Fig1}a shows a schematic of the experimental setup, where the sample is infused into the inlet of the microdevice using the withdraw function of the syringe pump located at the outlet. With this proposed system, the sample-containing Eppendorf can be gently stirred at regular time intervals to minimise cell sedimentation, which typically limits the duration of experiments. To test the effect of inlet conditions, inlet ports are manually punched onto the PDMS, so that the tubing connection can be either aligned (Fig.~\ref{fig:Fig1}b) or misaligned (Fig.~\ref{fig:Fig1}c) with the centreline of the microdevice. The results presented in this study are for the aligned case, unless otherwise stated.

The experimental setup used an inverted microscope  (Olympus IX71) for flow visualisation and a syringe pump (Nemesys, Cetoni) for fluid control. The microfluidic device was illuminated using a halogen lamp. Videos were acquired by a sensitive monochrome CCD camera (Olympus, XM10) using a $10\times$ magnification objective with numerical aperture of 0.25 to guarantee that the depth of field and image depth are large enough to observe cell trajectories across the entire channel height (Fig.~\ref{fig:Fig1}e and Sec.~\emph{\nameref{Data acquisition and image analysis, SI}}). An algorithm was developed \textit{in-house} based on ImageJ software to detect and count the cells automatically (Fig.~\ref{fig:FigS3} and Fig.~\ref{fig:FigS4} in Sec.~\emph{\nameref{Data acquisition and image analysis, SI}}). The RBC probability density distribution (PDD) along the width direction of a given region of interest (ROI, see Fig.~\ref{fig:Fig1}d) is then calculated using a minimum of 300 frames, containing more than 8,000 cells (Fig.~\ref{fig:FigS5} in Sec.~\emph{\nameref{Probability density distribution (PDD) of RBCs, SI}}). The cell-free layer (CFL) is determined as half the difference between the overall channel width and the RBC-core width, using composite images (see Fig.~\ref{fig:Fig1}f) generated via the Z-projection method. The detailed procedure is described Fig.~\ref{fig:FigS8} and Sec.~\emph{\nameref{Cell-free layer (CFL), SI}}.

\subsection{Computational modelling} 
\label{Computational modelling}

The immersed-boundary-lattice-Boltzmann method (IB-LBM \cite{kruger_computer_2012}) is employed to model blood flow as a suspension of deformable RBCs. The fluid flow governed by the Navier-Stokes equations is solved by the lattice-Boltzmann method with standard D3Q19 lattice \cite{qian_lattice_1992}, BGK collision operator \cite{bhatnagar_model_1954}, Guo's forcing scheme \cite{guo_discrete_2002} and the Bouzidi-Firdaouss-Lallemand (BFL) implementation of no-slip boundary condition at the walls \cite{bouzidi_momentum_2001}. To precisely control the volume flow rate within the channel, inflow/outflow open boundaries are implemented with the Ladd velocity boundary condition \cite{ladd_numerical_1994}. The RBCs are modelled as Lagrangian membranes using a finite element approach. The fluid flow and RBC membrane are coupled through the immersed-boundary method \cite{peskin_immersed_2002}, which tackles the interpolation of velocities and the spreading of forces. The model is implemented in the parallel lattice-Boltzmann flow simulator $HemeLB$ \cite{bernabeu_computer_2014} (open source at \url{http://ccs.chem.ucl.ac.uk/hemelb}) by incorporating a new module for discrete deformable cells. Details about the model configuration (Fig.~\ref{fig:FigS10}), simulation setup (Fig.~\ref{fig:FigS11}) and numerical analysis (Fig.~\ref{fig:FigS12}-\ref{fig:FigS14}) can be found in Sec.~\emph{\nameref{Model configuration, SI}}, Sec.~\emph{\nameref{Simulation setup, SI}} and Sec.~\emph{\nameref{Numerical analysis, SI}}. See Table \ref{table:T1} for the key simulation parameters.

Each RBC is modelled as a closed finite-element membrane with 720 triangular facets (meshing resolution matching the lattice size $\Delta x$ for numerical stability and accuracy \cite{kruger_efficient_2011}), with its unstressed shape assumed as a biconcave discoid. The RBC membrane is elastic, with its mechanical properties controlled by five moduli which govern different energy contributions. They are strain modulus $\kappa_{s}$, bending modulus $\kappa_{b}$, dilation modulus $\kappa_{\alpha}$, surface modulus $\kappa_{S}$ and volume modulus $\kappa_{V}$. For more details of the RBC model, please refer to \cite{kruger_computer_2012}. \REV{To tackle close contact among cells or between the cell and the wall (albeit rare in the dilute suspensions we consider), a repulsion potential decaying with inverse distance between neighbouring RBC surfaces is numerically implemented with interaction intensities (see $\kappa_{cc}$ and $\kappa_{cw}$ in Table \ref{table:T1}) comparable to the bending elasticity of the RBC membrane}. The behaviour of RBCs in flow is determined by two non-dimensional numbers: the particle Reynolds number $Re_\text{p}$ and the capillary number $Ca$. $Re_\text{p}$ represents the ratio of particle inertia to fluid viscous force and is defined based on the channel Reynolds number $Re_\text{c}$:
\begin{equation} 
\label{eq:1}
Re_\text{p} = Re_\text{c} \left(2r_\text{rbc}/D_\text{h}\right)^2, \quad Re_\text{c} = \rho_\text{ex} \bar{u} D_\text{h}/\eta_\text{ex}
\end{equation}
\noindent
where $r_\text{rbc}$ is the RBC radius and $D_\text{h}$ is the hydraulic diameter of the rectangular channel; $\bar{u}$ is the cross-sectional mean velocity of the unperturbed flow; $\rho_\text{ex}$ and $\eta_\text{ex}$ are the density and dynamic viscosity of the suspending fluid, respectively. $Ca$ represents shear-induced deformation of the RBC membrane and is defined using its shear elasticity (governed by the strain modulus $\kappa_{s}$):
\begin{equation} 
\label{eq:2}
Ca = \tau_\text{rbc}\dot{\gamma}_\text{c}, \quad \tau_\text{rbc} = \eta_\text{ex} r_\text{rbc}/\kappa_{s}, \quad \dot{\gamma}_\text{c} = 8 \bar{u}/D_\text{h}
\end{equation}
\noindent
where $\tau_\text{rbc}$ is a characteristic time scale for the RBC membrane to relax to its equilibrium shape from a transient state and $\dot{\gamma}_\text{c}$ is the characteristic wall shear rate in equivalent Poiseuille flow. Both $Ca$ and the viscosity contrast $\lambda$ (between RBC cytosol and the suspending fluid) are kept the same for experiments and simulations to ensure comparability (see values in Table \ref{table:T1}). 

\begin{table}[!t]
\caption{Key simulation parameters. The symbol ``$\sim$'' represents simulation units (dimensionless).}
\begin{tabular}{llr}
\hline
Parameter & Symbol & Value \\ 
\hline
\textit{channel geometry} & & \\ 
\hline
width & $W$ & 96 $\upmu$m \\
depth & $H$ & 30 $\upmu$m \\
length & $L$ & 1320 $\upmu$m \\
hydraulic diameter & $D_\text{h}$ & 45.7 $\upmu$m \\
\hline
\textit{RBC model} & & \\ 
\hline
cell radius & $r_\text{rbc}$ & 4 $\upmu$m \\
viscosity contrast & $\lambda$ & 1.0 \\
feeding haematocrit & $H_\text{F}$ & $1\%$ \\
local haematocrit & $Ht$ & variable \\
\hline
\textit{simulation setup} & & \\ 
\hline
particle Reynolds number & $Re_\text{p}$ & 0.03 \\
capillary number & $Ca$ & 0.6 \\
LB relaxation parameter & $\widetilde{\tau}_\text{BGK}$ & 1.0 \\
voxel size & $\Delta x$ & 0.6667 $\upmu$m \\ 
time step & $\Delta t$ & 7.41 $\times10^{-8}$s \\
shear modulus & $\widetilde\kappa_s$ & 5.00 $\times10^{-4}$ \\
bending modulus & $\widetilde\kappa_b$ & 4.50 $\times10^{-5}$ \\
reduced bending modulus & $\widetilde\kappa_b{\Delta x}^2/(\widetilde\kappa_s r_\text{rbc}^2$) & 1/400 \\
\REV{cell-cell interaction} & \REV{$\widetilde\kappa_{cc}$} & \REV{$\widetilde\kappa_b$} \\
\REV{cell-wall interaction} & \REV{$\widetilde\kappa_{cw}$} & \REV{10 $\widetilde\kappa_b$} \\
\hline
\end{tabular}
\label{table:T1}
\end{table}

\section{Results} 
\label{Results}

The experimental and simulation results on the RBC distribution in the channel are shown in Sec.~\emph{\nameref{Counterintuitive RBC distribution}}.
Additional data of RBC dynamics and the mechanisms leading to the observed RBC distribution is provided in Sec.~\emph{\nameref{Flow properties and RBC dynamics}}.

\subsection{Counterintuitive RBC distribution}
\label{Counterintuitive RBC distribution}

The main results of the study are reported here, focusing on the dynamical distribution of RBCs in a dilute suspension. Experimentally, by continuously feeding RBCs at fixed haematocrit ($H_\text{F} \leq$ 1\%) into the microfluidic device, the transverse motion of cells can be evaluated by monitoring the probability density distribution (PDD, see Sec.~\emph{\nameref{Probability density distribution (PDD) of RBCs, SI}}) of RBCs at sequential locations along the axial direction. This process is also replicated numerically to corroborate the experimental observations and provide further insights into the 3D motion of cells in a microchannel of rectangular cross-section.

\begin{figure*}[!t]
\centering
\includegraphics[width=1.0\linewidth]{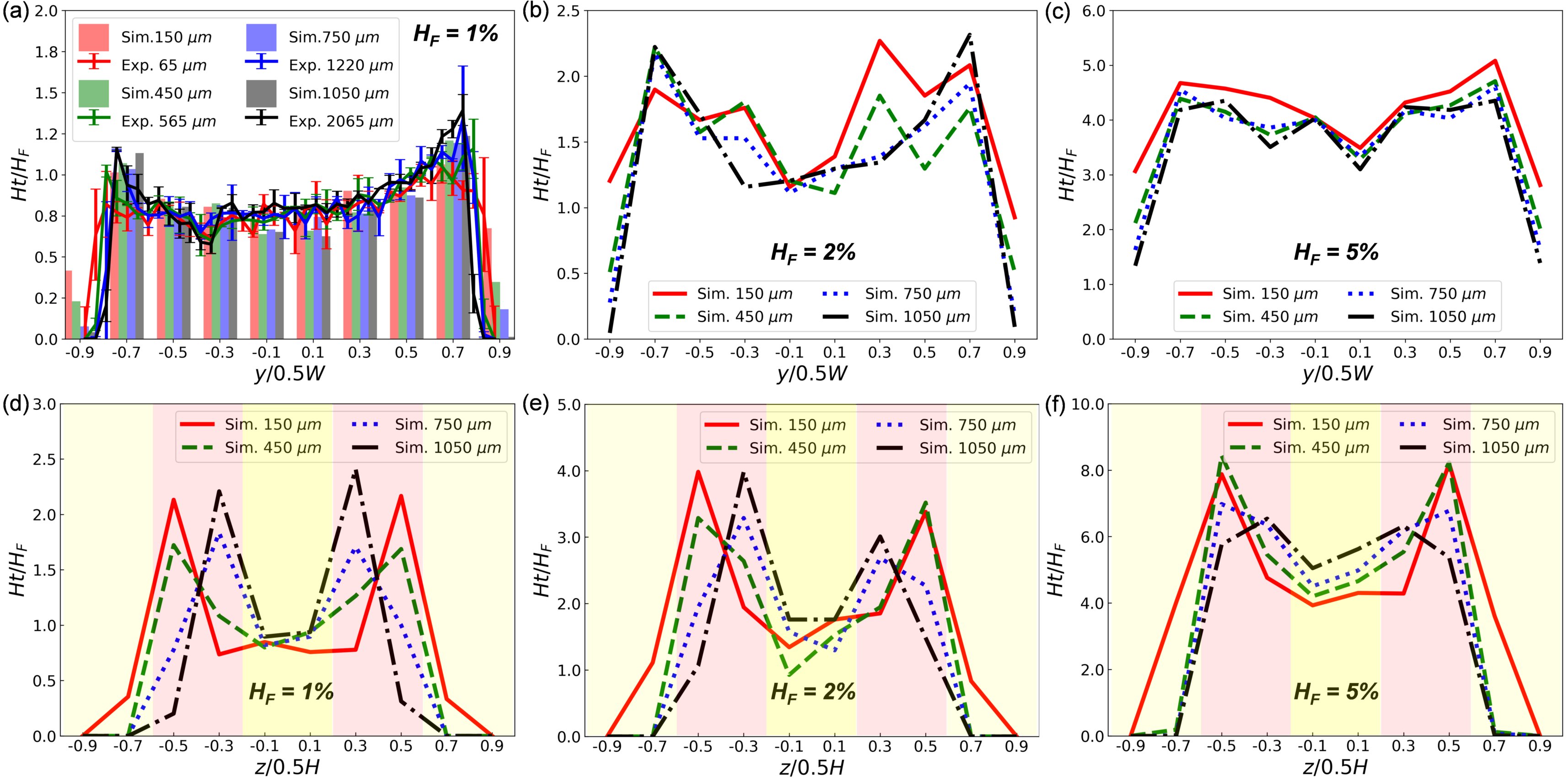}
\caption{\label{fig:Fig2} \REV{RBC distribution across the channel obtained by projecting cells either horizontally or vertically. (a) Experimental PDD profiles (lines) at $x = 65$, $565$, $1220$, $2065~\upmu$m versus simulation haematocrit $Ht$ histograms (bars) at $x = 150$, $450$, $750$, $1050~\upmu$m, showing the widthwise distribution of RBCs at $H_\text{F} =$ 1\%. Note that the experimental PDDs have been scaled by a constant to be superimposed onto the simulation $Ht$ histograms. (b) (Simulation) Widthwise $Ht$ profiles at $H_\text{F} =$ 2\%. (c) (Simulation) Widthwise $Ht$ profiles at $H_\text{F} =$ 5\%. (d--f) (Simulation) Corresponding depthwise $Ht$ profiles at $H_\text{F} =$ 1\%, 2\%, 5\%, respectively.}}
\end{figure*}

\paragraph{Off-centre two-peak (OCTP) profile}

In experiments with negligible inertia ($Re_\text{p} < 2 \cdot 10^{-4}$), we observe an evident increase of cell concentration close to each wall, manifested by the augmentation of two off-centre peaks in the PDD profile depicting RBC fractions across the channel width (Fig.~\ref{fig:Fig2}a, $Q = 0.2\,\upmu$L/min). Locations of the two peaks are approximately symmetric about the channel centreline and keep moving inwards until a certain distance about 0.2$\sim$0.3$W/2$ from the wall is reached (i.e., $y/(0.5W) = 0.7\sim0.8$). Beyond this distance, the inward migration of cells ceases (compare $x = 1220,\,2065\,\upmu$m in Fig.~\ref{fig:Fig2}a); this eventually causes an off-centre two-peak (OCTP) ordering in the density profile. Contrarily, in the central region, the cell concentration remains nearly unchanged. By comparing the PDD profiles along the channel length, a cell population originally close to each wall is found to cross streamlines and progressively shift inwards (see $x = 65,\,565,\,1220\,\upmu$m). \REV{For an initial cell distribution with significant left-right asymmetry induced by misaligned flow inlet (see Fig.~\ref{fig:Fig1}d), the same inward motion of cells near the wall occurs (Fig.~\ref{fig:FigS6}, SI)}. The observed quasi-steady OCTP distribution of RBCs clearly deviates from the commonly reported cell distribution (or haematocrit profile) in microscale channels, which typically exhibits a core of cells near the channel centre accompanied by a cell-depleted layer near the channel wall \cite{cokelet_decreased_1991, fedosov_blood_2010}. \REV{To rule out the possibility that such an OCTP distribution is exclusive to the volume flow rate $Q = 0.2\,\upmu$L/min, $Q$ is also varied up to 4.0 $\upmu$L/min for the aligned configuration (Fig.~\ref{fig:FigS7}, SI)}. Qualitatively similar evolution of the PDD profile is identified for all cases, featuring the development/enhancement of density peaks at two symmetric locations near the lateral channel walls.

To understand the peculiar cell distribution observed in experiments, numerical simulations for a $H_\text{F} =$ 1\% RBC suspension flowing at low Reynolds number ($Re_\text{p}$ = 0.03) are performed in a long straight microchannel of identical hydraulic diameter ($D_h=45.7\upmu$m) and similar aspect ratio ($AR = W/H=3.2$). In simulations, we have access to the three-dimensional organisation of RBCs within the channel and can observe the RBC distribution from not only the width direction $W$ but also the depth direction $H$, contrary to common experimental set-ups where only one dimension is observable. To quantify the RBC distribution, we analyse the time-averaged haematocrit ($Ht$) distribution across respective directions of the channel (Fig.~\ref{fig:Fig3}a). \REV{In the $W$ direction, the RBC distribution agrees well with experimental findings of the OCTP pattern and equivalent ``pseudo-equilibrium'' position of the peaks are found, i.e., $y_\text{eq}/(0.5W) = 0.7$ (Fig.~\ref{fig:Fig2}a)}. 

A distinct initial state of the suspension is found in the $H$ direction upon the entry of cells into the rectangular channel (at $x = 150\,\upmu\text{m} \approx 3.3 D_\text{h}$). While the widthwise distribution presents a more or less centralised pattern (\REV{Fig.~\ref{fig:Fig2}a}), the depthwise distribution features two primary density peaks (\REV{Fig.~\ref{fig:Fig2}d}). Furthermore, a substantial inward march of cells towards the channel centreline can be observed, with marked shifts of both density peaks. Downstream the channel, a pronounced five-layered ordering develops, featuring a cell-depletion band next to each wall, two cell-enrichment bands neighbouring the centre, and one cell-suppression band right at the centre (see $x = 1050\,\upmu\text{m} \approx 23 D_\text{h}$ in Fig.~\ref{fig:Fig2}d). An arguable equilibrium position for the primary density peaks exists at about 0.7 times the half channel depth away from the wall, i.e., $z_\text{eq}/(0.5H) = 0.3$.

\REV{

\begin{figure*}[!b]
\centering
\includegraphics[width=1.0\linewidth]{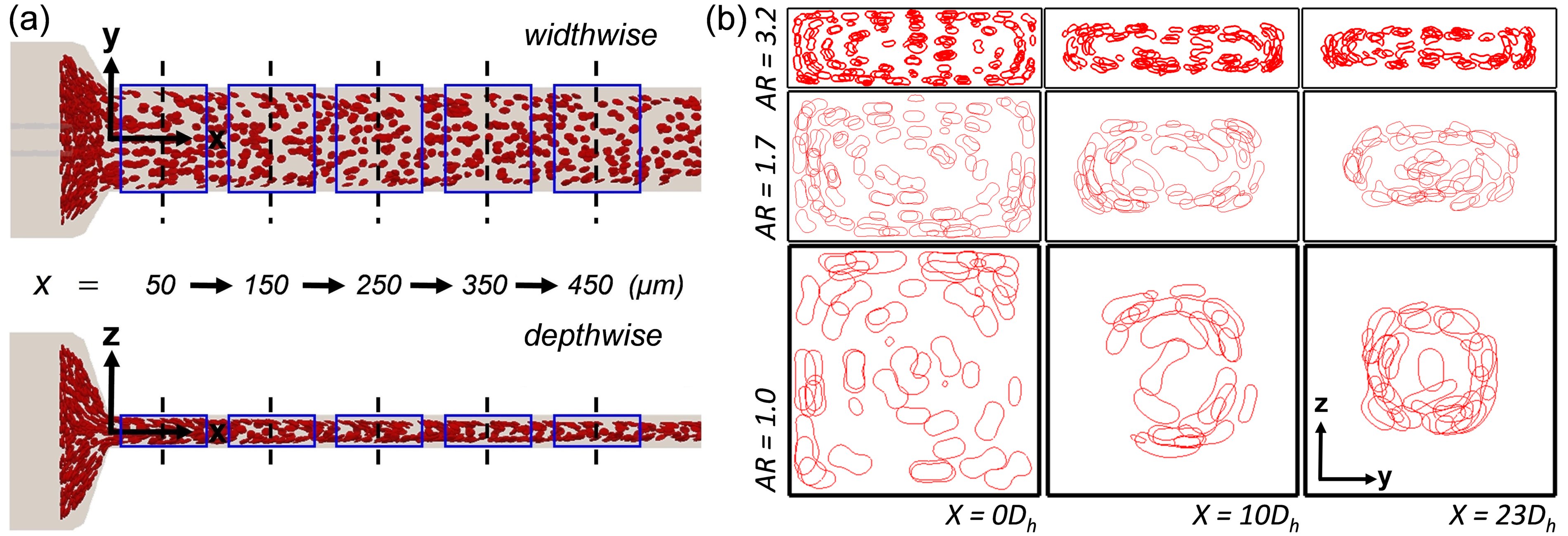}
\caption{\label{fig:Fig3} \REV{(Simulation) Cross-sectional visualization of RBC patterns. (a) Example of the haematocrit ($Ht$) analysis showing superimposed RBCs from multiple simulation snapshots (at designated time intervals) in sequential sampling boxes (blue) measuring $50~\upmu$m long each at target locations (ROIs) downstream of the channel entrance. Cells within each sampling box are projected and allocated into 10 bins across the width ($W$) and depth ($H$) directions for counting, respectively, with their position determined by centre of mass. (b) Cross-sectional slices showing the accumulative transverse pattern of RBCs at $x = 0D_\text{h}, 10D_\text{h}, 23D_\text{h}$ away from the entrance, combining snapshots from 50 consecutive time steps. The 1st, 2nd and 3rd row represent snapshots of the $H_\text{F} =$ 1\% RBC suspension in a straight rectangular channel with aspect ratios $AR = 3.2,\,1.7,1.0$ (varying $W = 96,\,50.5,\,30\,\upmu\text{m}$ while keeping $H = 30\,\upmu\text{m}$ fixed), respectively.}}
\end{figure*}

\paragraph{Effect of feeding haematocrit}

The OCTP pattern can still be observed in simulated RBC suspensions of moderately higher concentration, where the feeding haematocrit $H_\text{F}$ is increased to 2\% (Fig.~\ref{fig:Fig2}b,e) and 5\% (Fig.~\ref{fig:Fig2}c,f), respectively. However, the magnitude of the near-wall density peaks relative to the central region are in general weakened. Notably, in the depthwise distribution of $H_\text{F} =$ 5\% (Fig.~\ref{fig:Fig2}f), the two peaks (for which the local haematocrits can reach as high as 8-9 \% and presumably violating the dilute limit) are progressively smoothed out downstream the channel and the OCTP pattern vanishes at $x = 1050\,\upmu$m. Meanwhile, the central band is growingly enriched with cells and the over $Ht$ profile becomes flattened, suggesting considerable RBC dispersion taking place to even out regional density heterogeneities.

\paragraph{Effect of channel aspect ratio}

The occurrence of OCTP patterns across both dimensions of the channel suggests the phenomenon of RBC focusing. Indeed, by investigating the transverse organisation of RBCs in sequential cross-sections ($x = 0D_\text{h}, 10D_\text{h}, 23D_\text{h}$), we observe the gradual formation of an RBC loop reminiscent of the classic ``tubular pinch effect'' arising from inertia (Fig.~\ref{fig:Fig3}b), yet our experiments and simulations are all under negligible inertia as indicated by the low Reynolds number. Reducing the channel aspect ratio from the original $AR = 3.2$ to $AR = 1.7$ and $AR = 1.0$ does not remove the focusing behaviour, but brings the RBC loop closer to channel centre. Examination of the $Ht$ profiles in $AR = 1.7,\,1.0$ channels against the original $AR = 3.2$ channel reveals an inward shift of the OCTP equilibrium position in the width direction from $y_\text{eq}/(0.5W) = 0.7$ to $y_\text{eq}/(0.5W) = 0.5,\,0.3$ (Fig.~\ref{fig:FigS12}a,c in SI). In the case of $AR = 1.0$ where the two dimensions are identical, the equilibria from the $W$ and $H$ directions become equivalent, i.e., $y_\text{eq}/(0.5W) = z_\text{eq}/(0.5H) = 0.3$ (Fig.~\ref{fig:FigS12}c,d in SI).
}

\subsection{Flow properties and RBC dynamics}
\label{Flow properties and RBC dynamics}

To reveal the underlying physics behind the intriguing OCTP distribution of RBCs presented in Sec.~\emph{\nameref{Counterintuitive RBC distribution}}, we investigate several key aspects of the problem, including the origin of transverse cell migration, the effect of channel geometry, the distribution of cell velocity, the formation of the cell-depletion layer and the variation of tube haematocrit. The two main questions we aim to address are: 1) How do the two density peaks in the PDD profile come into being? 2) Why do the peaks keep building up instead of being dispersed?

\paragraph{Mechanism for cell migration}

Given its small Reynolds number, our RBC flow practically falls into the low-inertia regime, where viscous effects dominate. For deformable objects in viscous flow, the phenomenon of their lateral drift away from the wall is well-documented, dating back to the 1830s when Poiseuille first observed a plasma layer (or cell-free layer, CFL) in frog arterioles and venules \cite{sutera_history_1993}. The origin of such motion of RBCs was later determined to be the non-inertial hydrodynamic lift arising from unbalanced pressure forces on the cell. This lift can drive cells in microscale vessels/channels to cross streamlines and migrate towards the central area, ultimately leading to the formation of a two-phase flow featuring a cell core and a fluid periphery \cite{cokelet_decreased_1991, fedosov_blood_2010}. Following the discovery by Goldsmith and Mason \cite{goldsmith_axial_1961}, the lateral migration of deformed particles in non-inertial flows has been confirmed by other experimental studies \cite{lorz_weakly_2000, abkarian_tank_2002, callens_hydrodynamic_2008, coupier_noninertial_2008} and numerous modelling reports \cite{cantat_lift_1999, sukumaran_influence_2001, secomb_two-dimensional_2007, kaoui_lateral_2008, doddi_lateral_2008, meslinger_dynamical_2009, shi_lateral_2012, hariprasad_prediction_2015, farutin_analytical_2013}. In principle, there are three mechanisms for the cross-streamline motion of RBCs in low-Reynolds-number shear flows: (1) wall-repulsion-induced lift force on RBCs, (2) shear-gradient-induced asymmetric deformation of RBCs, and (3) hydrodynamic diffusion (or shear-induced diffusion) between RBCs due to cell-cell interactions. Depending on the flow properties (e.g., velocity profile, shear distribution and wall confinement) and cell conditions (e.g., position, shape and rigidity), the direction and strength of RBC motion under the combined effect of the above mechanisms can be difficult to determine.

\paragraph{Decay of lift velocity}

\begin{figure*}[!t]
\begin{minipage}{0.52\linewidth}
\includegraphics[width=1.0\linewidth]{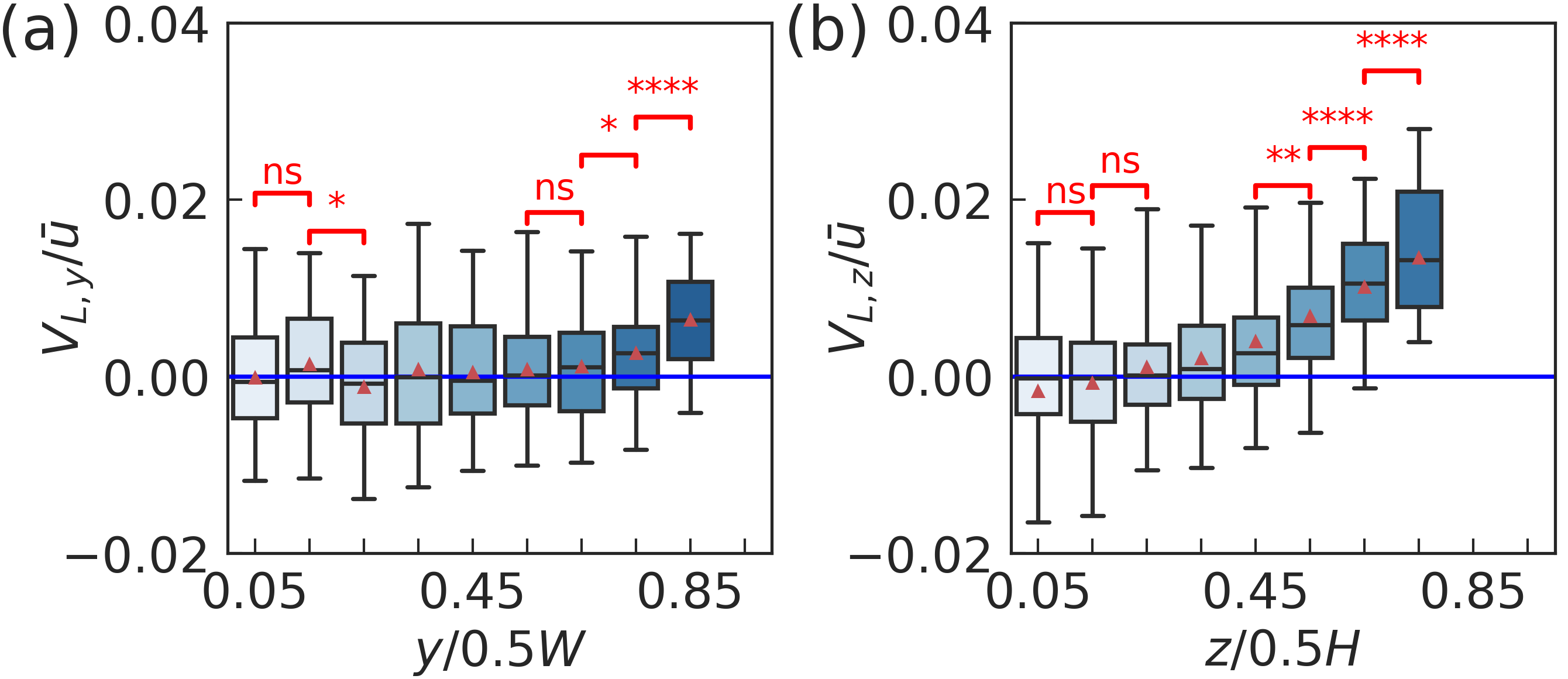}
\end{minipage}%%
\hfill
\begin{minipage}{0.46\linewidth}
\caption{\label{fig:Fig4} (Simulation) Statistical* lift velocities of RBCs within $x = (50\pm 8)\,\upmu$m, characterising the RBC migration along the (a) $y$-direction (widthwise) and (b) $z$-direction (depthwise), respectively. For statistical analysis in both (a) and (b), the channel is fictitiously folded along the channel centreline and then divided into 10 bins, followed by the assignment of cells according to the position of their center of mass. A positive lift velocity here indicates an inward motion of RBCs towards the channel centreline, vice versa}
\end{minipage}
\begin{minipage}{1.0\linewidth}
\textit{\footnotesize *The short line and triangle in the boxplots represent the median and mean lift velocity for each bin, respectively. The horizontal caps show the statistics from Welch's t-tests between neighbouring groups, where *P $<$ 0.05 is considered as statistical significance; **P $<$ 0.01, ****P $<$ 0.0001.}
\end{minipage}
\end{figure*}

The presence of RBCs at $H_\text{F} = 1\%$ (or below) does not significantly disturb the apparent velocity profiles from their unperturbed states (compare Fig.~\ref{fig:FigS15}c,d and Fig.~\ref{fig:FigS16} of the SI), we can therefore analyse the transverse motion of RBCs with respect to the unperturbed flow. Subject to the complex flow pattern in the channel as elaborated by Fig.~\ref{fig:FigS15}, the motion of RBCs within are likely to be affected by both the solid wall and the fluid shear gradients. At low Reynolds number, this motion can be conveniently decomposed into two orthogonal directions, hereby the $W$ and $H$ directions, on assuming negligible inertia and invoking the Stokes equations. Fig.~\ref{fig:Fig4}a--b show the statistical RBC lift velocities $V_\text{l}$ (relative to the unperturbed flow mean velocity $\bar{u}$) in 10 subdivisions of a cross-section, divided either widthwise ($W$) or depthwise ($H$). Compared with the depthwise lift, the widthwise lift is found to decay much faster as the cell-wall distance increases. While $V_{\text{l},z}$ vanishes at 0.75 times of $H/2$ away from the wall (i.e., $z/(0.5H) = 0.25$), $V_{\text{l},y}$ already becomes negligible at 0.35 times of $W/2$ away from the wall (i.e., $y/(0.5W) = 0.65$). The contrast of lift decay between the two directions can be explained by the shear-rate and shear-rate-gradient profiles in the cross-section (Fig.~\ref{fig:FigS15}e--f, SI). As the cells migrate towards the channel centreline, the local shear rate diminishes to zero, with much faster drop in $W$ than in $H$. Meanwhile, the local shear gradient also quickly vanishes in $W$ direction, whereas that in $H$ direction decreases in a relatively mild way. Consequently, contributions of both the wall-induced lift and shear-gradient-induced lift in the $W$ direction are weakened more quickly.

The distribution of lift velocities in both the width and depth directions coincides with the overall evolution of the $Ht$ profiles presented in Fig.~\ref{fig:Fig2}a,d: whereas a ``pseudo-equilibrium'' position close to the wall exists for the widthwise migration (beyond which the lateral motion of cells ceases), the depthwise cell migration is more substantial and results in a ``pseudo-equilibrium'' position much closer to the channel centre. Therefore, a correlation between the decaying lift of cells and their spatial distribution in the cross-sectional direction can be inferred.

\paragraph{Growth of cell-free layer}

\begin{figure*}[!t]
\centering
\includegraphics[width=0.7\linewidth]{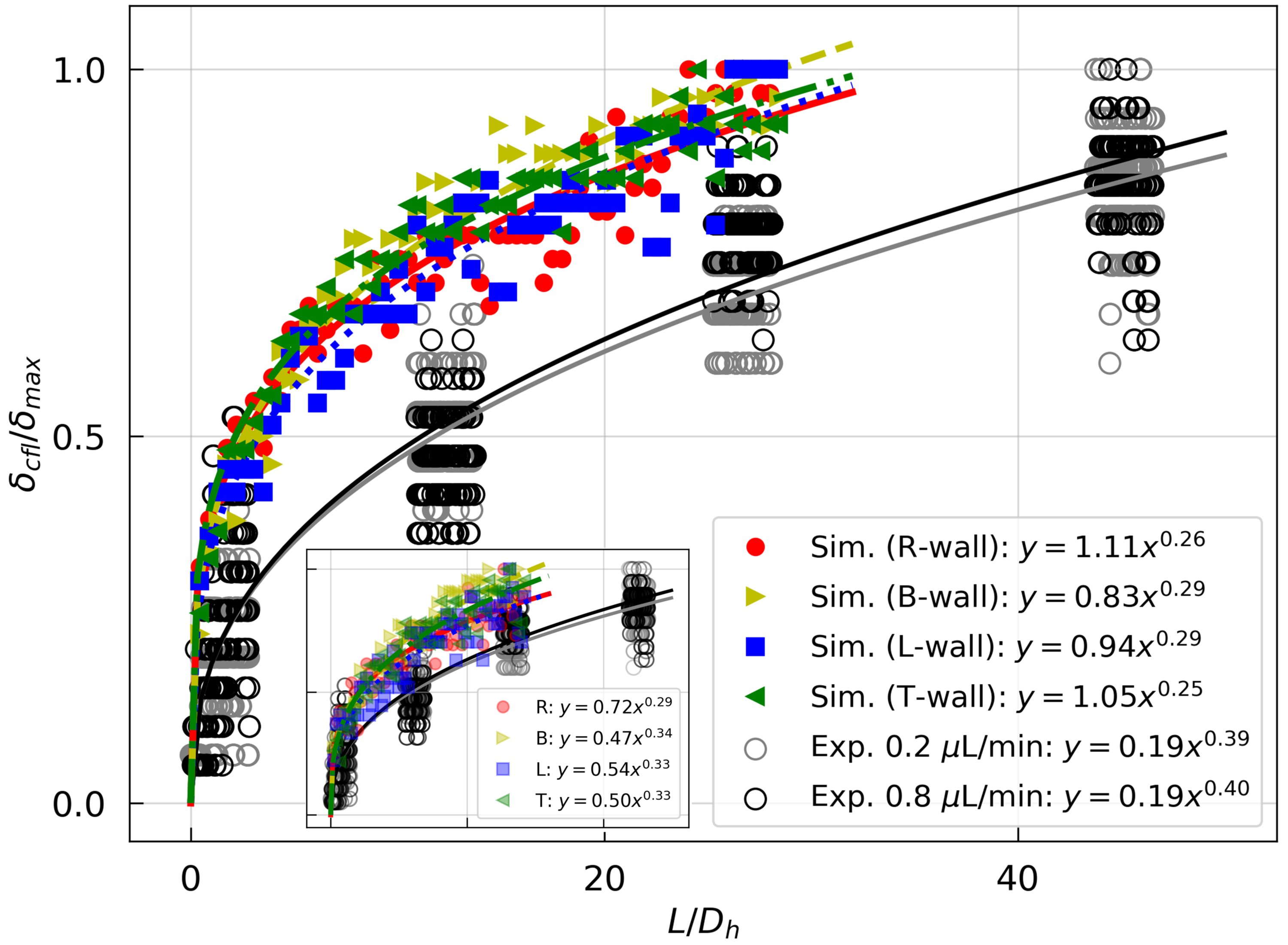}
\caption{\label{fig:Fig5} (Empty circles) Experimental CFL measured in the width direction of the channel (averaged between opposite walls) under $Q = 0.2\,\upmu$L/min and $Q = 0.8\,\upmu$L/min. (Solid symbols) Numerical CFLs calculated along each wall of the channel (Right/Bottom/Left/Top-walls). Both the experimental and numerical CFL values $\delta_\text{cfl}$ have been normalised by the maximum CFL thickness $\delta_\text{max}$ detected within the whole investigated range, and the development lengths $L$ are normalised by the channel hydraulic diameter $D_\text{h}$. The main frame and the inset of the figure contain the same experimental data, but different numerical data from two simulations: the former with normal human RBCs where $Ca = 0.6$ and the latter with ageing human RBCs where $Ca = 0.1$.}
\end{figure*}

We now move on to investigate how the transverse motion of RBCs contributes to the spatial development of CFL along the channel axis. In the lower-flow-rate experiment $Q = 0.2\,\upmu$L/min ($H_\text{F} \leq$ 1\%), the growth of CFL thickness (recording the averaged CFL from the two lateral walls) over $46D_\text{h}$ manifests an evident power-law behaviour with a best-fit exponent of 0.39 ($R^2 = 0.93$, Fig.~\ref{fig:Fig5}). \REV{For the other experiment with higher flow rate $Q = 0.8\,\upmu$L/min (hence larger $Re$), nearly identical power-law behaviour is found for the CFL development.} Similarly, the numerical CFLs over $28D_\text{h}$ (recording separate CFLs for individual lateral/vertical walls in simulation) also present power-law trends, but with certain discrepancy in the fitting exponents when compared to the experimental value of 0.39: for the widthwise CFLs at the right and left walls, the values are 0.26 ($R^2 = 0.91$) and 0.29 ($R^2 = 0.94$), respectively; for the depthwise CFLs at the bottom and top walls, they are 0.29 ($R^2 = 0.91$) and 0.25 ($R^2 = 0.95$), respectively (Fig.~\ref{fig:Fig5}). \REV{Part of the discrepancy in CFL growth rates between the experiment and simulation is attributed to the usage of different RBC types, namely horse RBC and human RBC ($Ca  = 0.6$ here), the latter of which has higher deformability in nature when exposed to shear. By tuning down the shear elasticity of the modelled cells in simulation ($Ca = 0.1$), which imitates ageing human RBCs of increased stiffness, the discrepancy between the experimental and numerical data is significantly reduced with numerical exponents of 0.29-0.34 achieved (inset of Fig.~\ref{fig:Fig5}).}

Despite the difference between the growth rates of experimental and simulation CFL, a neutral finding is that no critical saturation of the CFL increase can be identified throughout the investigated range \REV{(see raw data of the CFLs in Fig.~\ref{fig:FigS9}b and Fig.~\ref{fig:FigS13}a of the SI)}. This means that the development of CFL in a typical microfluidic channel (rectangular cross-section) for a dilute suspension can take much longer than earlier estimates provided for microvessels or cylindrical tubes \cite{pries_red_1989, katanov_microvascular_2015, ye_recovery_2016, ng_symmetry_2016}. In particular, Katanov et al. \cite{katanov_microvascular_2015} reported a universal length $L_\text{c} \leq 25D$ (independent of flow rate and haematocrit) for full CFL development in vessels between $D$ = 10$\sim$100$\,\upmu$m, whereas we find $L_\text{c} > 28D_\text{h}$ in simulation (only serving as a lower-bound estimate because we have limited our flow domain to such a length for tractable computational cost) and $L_\text{c} > 46D_\text{h}$ in experiment (hydraulic diameter of the channel $D_\text{h} \approx 45.7 \,\upmu$m). Based on our numerical findings of the ``pseudo-equilibrium'' positions for the widthwise cell migration (until 0.3 times of half channel width from wall) and depthwise cell migration (until 0.7 times of half channel depth from wall), the CFL thickness in the simulation may continue to grow until reaching 10$\sim$14 $\upmu$m, which are considerably larger than the values of 6$\sim$8 $\upmu$m we have monitored within $L_\text{c} \leq 28D_\text{h}$ \REV{(Fig.~\ref{fig:FigS13}a, SI)}.

\paragraph{Reduction of tube haematocrit}

The long-range development of CFL will inevitably affect the local rheological properties of the suspension. As shown in Table \ref{table:T2}, the tube haematocrit $Ht$ measured from either of our simulated geometries (CTRAC-L: large-contraction inlet, all simulation data and plots presented elsewhere in the paper are from this geometry; CTRAC-S: small-contraction inlet. See more details of CTRAC-L and CTRAC-S in Sec.~\emph{\nameref{Model configuration, SI}}) drops continuously along the channel axis compared to the feeding haematocrit $H_\text{F}$, even after $1050\,\upmu$m ($\approx 23D_h$) away from the entrance. This reduced volume fraction of blood flow in microchannels versus the discharge haematocrit ($H_\text{D} \approx H_\text{F}$) is known as the F{\aa}hr{\ae}us effect, where $Ht/H_\text{D} = \bar{U}_\text{B}/\bar{V}_\text{x} < 1$ ($\bar{U}_\text{B}$ and $\bar{V}_\text{x}$ are average velocities of the overall blood flow and the cells, respectively) \cite{fahraeus_suspension_1929, albrecht_fahraeus_1979}. It stems from the cross-streamline motion of RBCs towards the high-velocity region around channel centreline, which empowers the cells within the channel to travel faster on average than the ambient fluid in the streamwise direction. Owing to the fairly long CFL development length ($>28 D_\text{h}$ in simulation), the cross-sectional distribution of RBC fluxes and consequently the average ratio of cell-to-fluid velocities (Fig.~\ref{fig:FigS17} and Fig.~\ref{fig:FigS18} in Sec.~\emph{\nameref{RBC lateral migration and axial acceleration, SI}}) continues to change till the very end of the channel, leaving consistent measurement of tube haematocrits within the microchannel difficult.

\begin{table*}[!t]
\caption{(Simulation) Variation of tube haematocrit along the channel axis.}
\begin{tabular}{lcccccc}
\hline
Inlet type & $H_\text{F}$ & $Ht$ ($150\,\upmu$m) & $Ht$ ($450\,\upmu$m) & $Ht$ ($750\,\upmu$m) & $Ht$ ($1050\,\upmu$m) & $Ht$ ($1150\,\upmu$m) \\ 
\hline
CTRAC-L & 1\% & 0.810\% & 0.749\% & 0.703\% & 0.696\% & 0.685\% \\
CTRAC-S  & 1\% & 0.705\% & 0.661\% & 0.659\% & 0.622\% & 0.619\% \\
\hline  
\end{tabular}
\label{table:T2}
\end{table*}

\section{Discussion} 
\label{Discussion}

In the following we will discuss and compare our findings in Sec.~\emph{\nameref{Results}} with existing studies and interpretations, and elaborate on their implications for experimental studies of dilute blood flow.

\begin{figure*}[!t]
\centering
\includegraphics[width=1.0\linewidth]{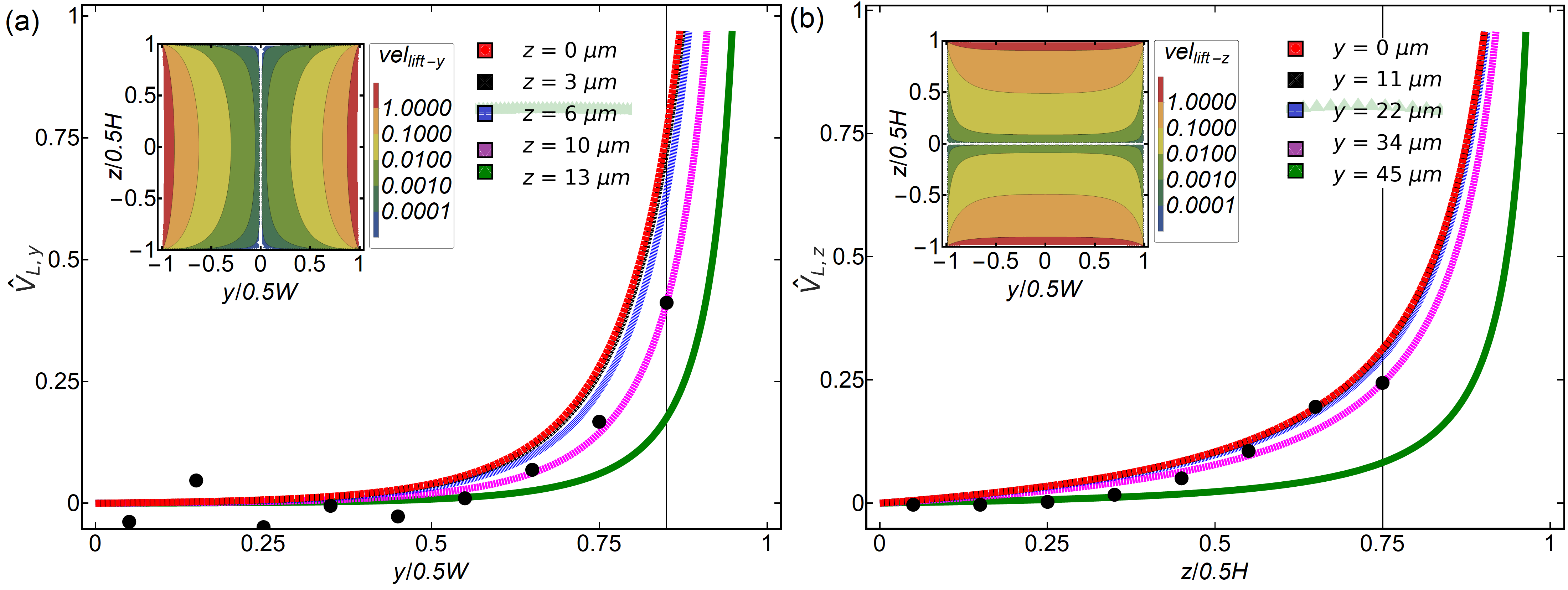}
\caption{\label{fig:Fig6} Theoretical prediction of RBC lift velocities (line curves and contour plots) given by $V_\text{l} \sim \dot{\gamma}/h$ across (a) the width direction and (b) the depth direction of the channel, respectively. $z$ and $y$ values in the legends represent positions relative to the mid-plane in each direction (see Fig.~\ref{fig:FigS15}c--d in the SI), with $z = 0$ and $y = 0$ denoting the widthwise and depthwise mid-planes, respectively. The black dots represent rescaled values of the numerical lift velocities extracted from Fig.~\ref{fig:Fig4}a--b. The black vertical lines in (a--b) indicate the center of the outmost layer of cells at $x = 50\,\upmu$m detected in the RBC simulation.}
\end{figure*}

\subsection{Comparison with existing studies}
\label{Comparison with existing studies}

\paragraph{Hydrodynamic lift}

The mechanism of hydrodynamic lift for tank-treading particles in shear flows has been analysed theoretically by Olla and Seifert in the 1990s \cite{olla_simplified_1999, seifert_hydrodynamic_1999}. A simplified scaling law $V_\text{l} \sim \dot{\gamma}/h^2$ was derived for the lift velocity $V_\text{l}$ under constant shear, as a function of the shear rate $\dot{\gamma}$ and cell-wall distance $h$ \cite{olla_lift_1997}. This quadratic scaling has been widely employed in the literature as an approximation to estimate the lateral migration of particles in various flow conditions. However, recent studies show that in pressure-driven channel flows where non-constant shear (or non-zero shear gradient) exists, e.g., Poiseuille flow, the curvature of the velocity profile modifies the scaling law in a non-trivial way and would significantly slow down the spatial decay of the lift velocity \cite{kaoui_lateral_2008, danker_vesicles_2009, farutin_analytical_2013, nix_lateral_2016}. Indeed, through systematic measurement of the lateral migration of various vesicles in bounded planar Poiseuille flow under different confinement degrees, Coupier et al. revealed that the scaling $V_\text{l} \sim \dot{\gamma}/h^2$ would fail to predict the trajectory of the vesicle, while a markedly different empirical law $V_\text{l} \sim \dot{\gamma}/h$ held \cite{coupier_noninertial_2008}. Losserand et al. confirmed the scaling $V_\text{l} \sim \dot{\gamma}/h$ using RBC experiments and further refined the empirical law \cite{losserand_migration_2019}.

To examine whether the lateral migration of RBCs in our simulation obeys the lift scaling that Coupier et al. \cite{coupier_noninertial_2008} and Losserand et al. \cite{losserand_migration_2019} proposed for Poiseuille flow, we adopt theoretical shear rates from the asymptotic solution of pressure-driven flow in rectangular channel \cite{bruus_theoretical_2008} and calculate the lift velocities applying $V_\text{l} \sim \dot{\gamma}/h$ (abandoning all prefactors for generality). Further with weighted normalisation, we obtain the dimensionless theoretical lifts ($\hat{V}_{\text{l},y}$ and $\hat{V}_{\text{l},z}$) for both $W$ and $H$ directions (Fig.~\ref{fig:Fig6}a--b). For comparison with the theoretical prediction, numerical medians of our numerical lift velocities are extracted from Fig.~\ref{fig:Fig4}a--b. After rescaling and overlapping them onto the theoretical curves, we observe satisfactory data agreement across the $H$ direction but only partial agreement in the $W$ direction for the near-wall region only. Approaching the central region in the $W$ direction, the theoretical prediction breaks down. This is actually expected since the velocity profile there is abruptly blunted and significantly deviates from a parabolic shape. We have also examined against our simulation data the theoretical prediction given by $V_\text{l} \sim \dot{\gamma}/h^2$, which would overestimate the spatial decay of the lift in general and failed to yield an agreement with the simulation data (results not shown here).

\paragraph{Cell-free layer}

CFL characterisation is an active research field owing to its prominent role in regulating blood viscosity and local haemodynamics. Limited by the complexity of in vivo imaging and measurement \cite{kim_cell-free_2009}, glass capillaries and PDMS microchannels have been widely used to mimic simplified micro-circulatory environment for in vitro observation of CFLs under different rheological conditions \cite{tripathi_passive_2015}. Meanwhile, versatile computational tools are developed and plenty of in silico studies on CFL emerge \cite{fedosov_blood_2010}. However, most in vitro/in silico studies assume a steady CFL for a given channel segment and hence fail to capture the highly dynamical nature of realistic CFLs as in the microcirculation \cite{kim_temporal_2007,ong_spatio-temporal_2012}; albeit some authors made improvements by proposing a nominal CFL development length by multiplying the mean flow velocity and characteristic flow time \cite{zhang_effect_2011,katanov_microvascular_2015}. It is only recently that exact analyses of the spatio-temporal heterogeneity of CFL were presented \cite{oulaid_cell-free_2014,ye_recovery_2016,balogh_cell-free_2019}, detailing the spatial development/recovery of the CFL alongside microchannels in the presence of upstream disturbances.

\begin{figure*}[!t]
\begin{minipage}{0.60\linewidth}
\includegraphics[width=1.0\linewidth]{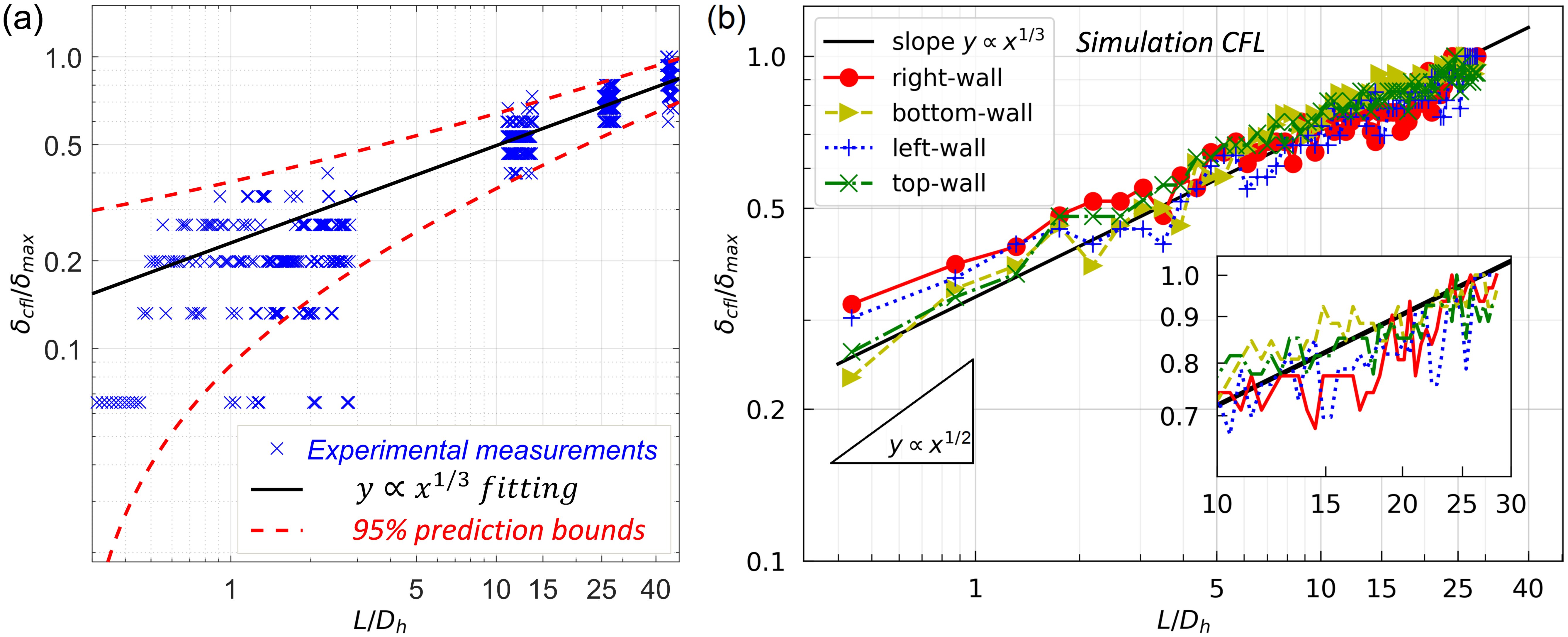}
\end{minipage}%%
\hfill
\begin{minipage}{0.39\linewidth}
\caption{\label{fig:Fig7} Regression analysis (log-log scale) of the (a) experimental and (b) numerical CFLs as in Fig.~\ref{fig:Fig5}. The CFL values $\delta_\text{cfl}$ are normalised by the maximum CFL thickness $\delta_\text{max}$ detected within the whole investigated range, and the development lengths $L$ are normalised by the channel hydraulic diameter $D_\text{h}$. The inset of (b) shows a zoom-in view of the numerical CFL trends within $L = 10\sim28 D_h$.}
\end{minipage}
\end{figure*}

In the present study, having confirmed the spatial variation of lift velocity $V_\text{l} \sim \dot{\gamma}/h$ (rather than $V_\text{l} \sim \dot{\gamma}/h^2$) against the cell-wall distance $h$ for individual RBCs, estimation can be made for the longitudinal development of $\delta_\text{cfl}$ in a dilute suspension using an analytical approach (see Sec.~\emph{\nameref{Scaling analysis, SI}} for further details). Our derivation shows that considering $a = 1$ (corresponding to $V_\text{l} \sim \dot{\gamma}/h$) in Eq.~\ref{eq:3} yields a power-law scaling for the CFL as expressed in Eq.~\ref{eq:13}: $\delta_\text{cfl} \sim \ell^\beta,\,\beta = 1/3$, where $\ell$ is the axial distance along the channel. The exponent of 1/3 here is surprising as it coincides with the power-law scaling for vesicle/RBC motion in simple shear flow with constant shear rate (see Eq.~3 in reference \cite{callens_hydrodynamic_2008}). Note that because $\delta_\text{cfl} \sim \ell^{1/3}$ is derived on the basis of a parabolic velocity profile under the assumption of planar Poiseuille flow (constant shear-rate gradient), whereas the flow patterns in a real microchannel are far more complex (see Fig.~\ref{fig:FigS15}, SI), its accuracy for the prediction of spatial CFL development remains to be verified by our experimental measurements and numerical results.

The power-law growth $\delta_\text{cfl} \sim \ell^{1/3}$ predicted by the scaling analysis assuming ideal Poiseuille flow turns out to agree well with both our experimental and numerical CFLs at $H_\text{F} =$ 1\%, for which the best-fitted power-law exponents were earlier shown to be 0.39 and 0.25$\sim$0.29, respectively (Fig.~\ref{fig:Fig5}). In further regression analyses of the experimental and numerical CFLs, both data sets are found to collapse on $\delta_\text{cfl} \sim \ell^{1/3}$ (Fig.~\ref{fig:Fig7}a--b). Furthermore, the instantaneous slopes for the numerical data are nearly independent of $\ell$ at the early stage of CFL development, presenting a roughly steady increase of 1/3 (see Fig.~\ref{fig:Fig7}b). This implies the predominant role of a single mechanism for the transverse motion of cells, which is sensibly the hydrodynamic lift as our scaling analysis suggests. Moving downstream of the channel, a continual increase of the CFL is observed upon approaching $\ell = 28 D_\text{h}$, but with slower growth indicated by decreasing slopes noticeably smaller than the original 1/3 (see inset of Fig.~\ref{fig:Fig7}b). This deceleration in CFL growth is expected down to the fact that the lift forces are weakened as the cells migrate further away from the wall and enter the low-shear region with blunted velocity profiles, where the lateral motion of cells slows down. \REV{In simulations of higher feeding haematocrit $H_\text{F} =$ 2\% and $H_\text{F} =$ 5\%, the scaling of $\delta_\text{cfl} \sim \ell^{1/3}$ still holds for early-stage development of the CFL until reaching far downstream of the channel (see Fig.~\ref{fig:FigS14}e and Fig.~\ref{fig:FigS14}f). Notably for $H_\text{F} =$ 5\%, the slope approximates zero between $\ell = 20-25 D_\text{h}$ and suggests a saturated CFL. This is in line with what was found in the evolution of $Ht$ profiles (Fig.~\ref{fig:Fig2}c,f), where exceedingly amplified local haematocrits (about 8-9\%) break the dilute regime and lead to considerable RBC dispersion that resists further drift of cells away from the wall.} 

From previous simulations of much denser RBC suspensions ($Ht =$15\%$\sim$45\%) in cylindrical tubes \cite{katanov_microvascular_2015}, a similar power-law behaviour was reported for the steep increase of CFL thickness at an initial development stage, before a second time scale induced by relaxation of the RBC core came into play. However, the power-law exponents under several shear rates were considerably smaller than 1/3 even at the very beginning. The deviation from 1/3 was caused by the cell-cell interactions existing in their dense suspensions, which obstructed the initial migration of RBCs and led to slower CFL growth than in a dilute suspension like ours. Arguably, the authors reasoned for the ideal growth factor of 1/3 by invoking the lift force scaling $F_\text{l} \sim V_\text{l} \sim 1/h^2$ (where $\dot{\gamma}$ is a constant and does not affect the scaling) derived from simple shear flows \cite{sukumaran_influence_2001, meslinger_dynamical_2009}. Albeit such reasoning does explain the qualitative trend of their CFL growth, the simple shear flow approximation itself for quadratic flows in a microchannel (where $\dot{\gamma}$ varies) is questionable, which may fail to capture important features of the RBC distribution in the presence of shear rate gradients as recent theories of particle migration at low Reynolds number demonstrate \cite{qi_theory_2017, henriquez_rivera_mechanistic_2016}.

\subsection{Implications and limitations}
\label{Implications and limitations} 

What we have covered so far not only elucidates the ``aberrant'' RBC distribution observed in our experiment and simulation contradicting earlier findings, but also has wider implications for experimentalists working on dilute suspensions, especially in high-aspect-ratio microfluidic devices. In brief, microfluidic designs need to be longer if their purpose is to investigate the microscopic behaviour of a dilute suspension \emph{after} completed lateral migration. Experimentally, it is notoriously difficult to accurately control the inlet distribution of a cell/particle suspension under various \emph{entrance effects}. Therefore, local observations are frequently reported based on the assumption that the suspension organisation is free of transient effects at the chosen region of interest (ROI) as long as the ROI is adequately far away from the entrance. However, the required length as revealed by our study turns out more demanding than commonly believed \cite{katanov_microvascular_2015, pries_red_1989, ye_recovery_2016, ng_symmetry_2016}. Some key messages are: 

(I) For experiments of dilute RBC suspensions in \emph{high-aspect-ratio} channels, e.g., $AR = W/H > 3$, the hydrodynamic lift of RBCs are substantially suppressed due to the existence of low-shear-rate and low-shear-gradient zones in the centre. Consequently, the full development of a CFL in these channels may require a much longer length ($L_\text{c} > 28D_\text{h}$ in simulations and $L_\text{c} > 46D_\text{h}$ in experiments) than what is normally expected for cylindrical tubes ($L_\text{c} \leq 25D$ according to reference \cite{katanov_microvascular_2015}). \REV{To achieve full CFL development within $L_\text{c} = 20-25D_\text{h}$ in rectangular channels, RBC suspensions with $H_\text{F} \geq 5$\% should be adopted}. Furthermore, the relevance of in vitro measurements of cellular behaviour using such high-aspect-ratio microchannels to the realistic microcirculatory blood flow needs reappraisal. After all, the vessels in the microcirculation are more likely to have circular cross-sections and considerably different velocity profiles, which the radial distribution of RBCs relies on. 

(II) When determining the organisation of a dilute suspension (e.g., haematocrit profile) in a typical microchannel of rectangular cross-section, local measurements may be misleading if the channel is short or only moderately long. This is because heterogeneities emerging in the suspension (e.g., density peaks) will persist and hinder the formation of an equilibrium cell distribution owing to the lack of RBC dispersion under \emph{weak cell-cell interactions}. In particular, if certain upstream disturbance exists (e.g., geometric constriction or expansion), history effects need to be considered when recording local suspension properties such as the tube haematocrit. Comparing the cases CTRAC-S and CTRAC-L in Table \ref{table:T2}, with channel geometry, unperturbed flow rate and feeding haematocrit all being the same, yet the tube haematocrits measured at sequential locations constantly differ by more than 6\% (relative change). The reason is simply their differently configured flow inlets, causing unequal degrees of disturbance to the initial distribution of cells. 

(III) Phenomena violating empirical laws may occur downstream of bifurcations if the upstream RBC suspension is at an intermediate stage of development and has special cell distribution, e.g., the OCTP profile observed in our study. Recently, counterintuitive inversion of haematocrit partition against the classic Zweifach-Fung effect in a bifurcating channel was reported by Shen et al. \cite{shen_inversion_2016}, arising from the formation of consecutive layers of high and low RBC haematocrits in the parent branch. The juxtaposed configuration of RBCs discovered by that study is in line with the five-layered ordering of cells we find in the depth direction of our RBC flow (Fig.~\ref{fig:Fig2}d--f). Therefore, the phenomenon Shen et al. captured might also have been a consequence of inadequate channel length for the suspension to be fully developed in the parent branch.

Some limitations of the present study are: 
(1) While the numerical model simulates human RBCs, horse RBCs are used in experiments due to laboratory restrictions. \REV{This causes a certain degree of discrepancy between the simulation and experiment, but its influence on the suspensions behaviour we studied, such as cell focusing and CFL development, is shown to be quantitative instead of qualitative.} 
(2) The effect of inlet configuration on suspension organisation has not been characterised in a quantitative manner, which can potentially provide more specific guidance for experimentalists to improve their microfluidic designs by reducing transient effects. We aim to address this issue in future studies, involving geometric factors such as constrictions, expansions and bifurcations. 
(3) The experimental data is insufficient for analysis of instantaneous growth rates of the CFL thickness as has been performed for the simulation. More measurements using a denser arrangement of ROIs, though quite challenging, should be made in future experiments in order to enable precise comparison with simulations. 
(4) \REV{To make the real-scale (channel length > 1 mm) simulations tractable, we have largely reduced the parameter space and refrained from an exhaustive investigation combining a range of flow scenarios and cell conditions}. A parametric study integrating all these effects will lead to enhanced understanding of the intricate behaviour of dilute suspensions processed in microfluidic devices.

\section{Conclusion} 
\label{Conclusion}

We performed both microfluidic experiments and numerical simulations of dilute blood suspensions in a low-Reynolds-number channel flow ($Re_\text{p} \ll 1$). Surprisingly, an off-centre two-peak (OCTP) ordering of RBCs was found, which is reminiscent of the ``tubular pinch effect'' typical of the radial distribution of particles in inertial microfluidics \cite{segre_radial_1961, park_continuous_2009}. However, the transverse motion of cells in our case has an entirely different origin: the non-inertial hydrodynamic lift of deformable particles \cite{goldsmith_axial_1961}. 

Back to the two primary questions we set out to address (see the opening paragraph of Sec.~\emph{\nameref{Flow properties and RBC dynamics}}): how do the two density peaks in the PDD profile come into being and why do the peaks keep building up instead of being dispersed? The reason behind the persistence of significant density peaks is the deficiency of hydrodynamic interactions among cells in the dilute limit, where shear-induced diffusion remains weak and falls short to smooth out density heterogeneities brought by the lateral migration of cells. Consequently, the evolution of the density profiles is predominantly determined by the decay of hydrodynamic lift within the suspension, which exhibits distinct patterns in the width and depth directions of a high-aspect-ratio microchannel and contributes to a counterintuitive (OCTP) profile of cells in the larger dimension. Additionally, depending on the inflow configuration, the initial distribution of cells upon their entry into the channel varies substantially, which brings extra complexity to the restoration of a converged density profile and the development of an equilibrium cell free layer. 

Our findings highlight the importance of local and transient characteristics at the dilute limit for an RBC-laden flow in microchannels. Experimentalists should therefore be cautious when working with dilute suspensions in microfluidics and make judicious assumptions about suspension properties with the presence of upstream disturbance to the flow, especially when the microscopic behaviour of the suspension is the focus of research.

\paragraph{Author contributions}
QZ and JF designed the research, analysed the data and wrote the article. QZ performed the simulations. JF performed the experiments. LC assisted in validation of the computational model. MOB and PRH discussed the results and revised the manuscript. MOB supervised the study. MSNO and TK conceived the research idea, supervised the study and edited the manuscript. 

\paragraph{Acknowledgements}
QZ thanks the University of Edinburgh for the award of a Principal's Career Development Scholarship and an Edinburgh Global Research Scholarship. TK's and MOB's contributions have been funded through two Chancellor's Fellowships at the University of Edinburgh. MOB is supported by grants from EPSRC (EP/R029598/1, EP/R021600/1), Fondation Leducq (17 CVD 03), and the European Union’s Horizon 2020 research and innovation programme under grant agreement No 801423. Supercomputing time on the ARCHER UK National Supercomputing Service (http://www.archer.ac.uk) was provided by the ``UK Consortium on Mesoscale Engineering Sciences (UKCOMES)'' under the EPSRC Grant No. EP/R029598/1. The authors declare that they have no competing financial interests.

% \bibliography{RBCdynamics_clean}

\newpage
\appendix
\setcounter{page}{0}
\setcounter{figure}{0}
\setcounter{equation}{0}

\renewcommand{\thepage}{S\arabic{page}} 
\renewcommand{\thesection}{S\arabic{section}}  
\renewcommand{\thetable}{S\arabic{table}}  
\renewcommand{\thefigure}{S\arabic{figure}}
\renewcommand{\theequation}{S\arabic{equation}}

\begin{center}
{\huge \textbf{Spatio-temporal dynamics of dilute red blood cell suspensions in low-inertia microchannel flow}} \\
\bigskip
{\huge Supplemental Information (SI)}

\bigskip
\textit{Qi $Zhou^{1,+}$, Joana $Fidalgo^{2,+}$, Lavinia $Calvi^{1}$, Miguel O. $Bernabeu^{3}$, Peter R. $Hoskins^{4}$, M\'onica S. N. Oliveir$a^{2, *}$, Timm Kr\"uge$r^{1,*}$}

\bigskip
\textit{1. School of Engineering, Institute for Multiscale Thermofluids, University of Edinburgh, Edinburgh EH9
3FB, UK}

\textit{2. James Weir Fluids Laboratory, Department of Mechanical and Aerospace Engineering, University of
Strathclyde, Glasgow G1 1XJ, UK}

\textit{3. Centre for Medical Informatics, Usher Institute, University of Edinburgh, Edinburgh EH16 4UX, UK}

\textit{4. Centre for Cardiovascular Science, University of Edinburgh, Edinburgh EH16 4SB, UK}

\textit{+ These authors contribute equally to this work.} \\
\textit{* Correspondence: monica.oliveira@strath.ac.uk or timm.krueger@ed.ac.uk}
\end{center}

\newpage
\section{Scaling analysis}
\label{Scaling analysis, SI}

\paragraph{RBC migration} 
Consider an RBC whose center of mass is initially located at a position $(x_0,y_0)$ in close proximity to the wall ($y = 0$) upon entry into a microchannel of hydraulic radius $R_\text{h}$. To make the asymptotic analysis tractable, we consider a unidirectional flow along the axial direction of the channel with velocity $u_\text{x}(y)$, which is a function of the distance to the channel wall. Driven by the flow, the RBC will gradually migrate towards the channel centreline ($y = R_\text{h}$) under hydrodynamic lift while travelling along the streamwise direction (i.e., $x$-direction), for which the motion can be described by a two-component velocity $v(v_\text{x},v_\text{y})$. Note that $v$ is deliberately used here to represent cell velocity while $u$ is reserved for fluid velocity. For leading-order estimation, we assume $v_\text{x} := u_\text{x}(y)$. Based on the empirical law by Losserand et. al (1), we propose an expression for the lift velocity of the RBC $v_\text{y}$ as
\begin{equation} 
\label{eq:3}
v_\text{y} = \dot{y} = K\dot{\gamma}(y)y^{-a},\quad \dot{\gamma}(y) = \frac{\partial u_\text{x}(y)}{\partial y},
\end{equation}
where $\dot{\gamma}$ is the shear rate of the unperturbed flow, and $K$ and $a$ are constants depending on flow conditions and cell properties. For a given period of time, e.g., $t\in[0,t']$, the RBC travels to a new position $(x',y')$, and the axial distance $\ell$ can be written as
\begin{equation} 
\label{eq:4}
\ell = x'-x_0 = \int_{0}^{t'}v_\text{x}~\text{d}t = \int_{y_0}^{y'}u_\text{x}(y)~\frac{\text{d}t}{\text{d}y}{\text{d}y} = \int_{y_0}^{y'}\frac{u_\text{x}(y)}{\dot{y}}~{\text{d}y}.
\end{equation}
Combining Eqs.~(\ref{eq:3}) and (\ref{eq:4}), one finds:
\begin{equation} 
\label{eq:5}
\ell = \frac{1}{K}\int_{y_0}^{y'}y^{a}v_\text{x}(y)\left(\frac{\partial u_\text{x}(y)}{\partial y}\right)^{-1}~{\text{d}y}.
\end{equation}
On assuming Poiseuille flow, the form of $u_\text{x}$ is known and $v_\text{x}$ may be expressed as
\begin{equation} 
\label{eq:6}
v_\text{x} = u_\text{x}(y) = \hat{u_\text{x}}\left(1-\left(\frac{R_h-y}{R_h}\right)^2\right),\quad \hat{u_\text{x}} = 2\bar{u},
\end{equation}
where $\hat{u_\text{x}}$ and $\bar{u}$ are the maximum and mean flow velocity, respectively. By replacing $v_\text{x}$ in Eq.~\ref{eq:5} with the above, one gets:
\begin{equation} 
\label{eq:7}
\ell = \frac{1}{2K}\int_{y_0}^{y'}y^{a+1}\left(2+\frac{y}{R_h-y}\right)~{\text{d}y}.
\end{equation}
Given that lateral migration of the RBC away from the wall happens very slowly, we may further assume that $y \ll R_h$ over the time scale of our investigation. This means $y/(R_h-y)$ is small, which allows us to arrive at a straightforward expression of the axial distance $\ell$ that can be integrated over the lateral displacement:
\begin{equation} 
\label{eq:8}
\ell \approx \frac{1}{K}\int_{y_0}^{y'}y^{a+1}~{\text{d}y} = \left.\frac{y^{a+2}}{(a+2)K}\right\rvert_{y_0}^{y'}.
\end{equation}
After some rearrangement, we may write the equation as
\begin{equation} 
\label{eq:9}
y' = \left((a+2)K\ell+y_0^{a+2}\right)^{\frac{1}{a+2}}.
\end{equation}

\paragraph{CFL development} 
Now let us consider an arbitrary RBC cloud in the microchannel flow setting off from the wall. It is actually the outermost cells of the cloud that determine the CFL thickness $\delta_\text{cfl}$, representing the distance between the RBC membrane and the wall. At high $Ca$ numbers (e.g., $Ca = 0.6$), the RBCs are highly deformable and significantly elongated along the flow, and $\delta_\text{cfl}$ can be approximated by $y$, namely the distance of the RBC's centre of mass to the wall:
\begin{equation} 
\label{eq:10}
\delta_\text{cfl}(0) \approx y_0,\quad \delta_\text{cfl}(t') \approx y'.
\end{equation}
Combining Eqs.~(\ref{eq:9}) and (\ref{eq:10}), we arrive at an expression describing the growth of $\delta_\text{cfl}$ along the channel axis at sequential positions of $\ell$:
\begin{equation} 
\label{eq:11}
\delta_\text{cfl}(t') = \left((a+2)K\ell+\delta_\text{cfl}(0)^{a+2}\right)^{\frac{1}{a+2}}.
\end{equation}
Ideally, if the outermost cells are directly in contact with the wall, $\delta_\text{cfl}(0) = 0$. This simplification results in a power-law scaling between $\delta_\text{cfl}$ and $\ell$, with an exponent of $\beta$:
\begin{equation} 
\label{eq:12}
\delta_\text{cfl} \sim \ell^\beta,\quad \beta = \frac{1}{a+2}.
\end{equation}
Specifically, for $a = 1$, corresponding to $V_\text{l} \sim \dot{\gamma}/h$, the exponent becomes 1/3 and the below scaling holds:
\begin{equation} 
\label{eq:13}
\delta_\text{cfl} \sim \ell^{1/3}.
\end{equation}

\section{Experiments}
\label{Experiments, SI}

\subsection{Preparation of RBC suspensions} 
\label{Preparation of RBC suspensions, SI}

\begin{figure}[!h]
\begin{minipage}{0.40\linewidth}
\caption{\label{fig:FigS1} Photomicrography of the horse RBC suspension in PS (NaCl 0.9\% w/v) at different magnifications. (a) 40x, scale bar 10 $\upmu$m; (b) 10x, scale bar 50 $\upmu$m.}
\end{minipage}%%
\hfill
\begin{minipage}{0.58\linewidth}
\includegraphics[width=1.0\linewidth]{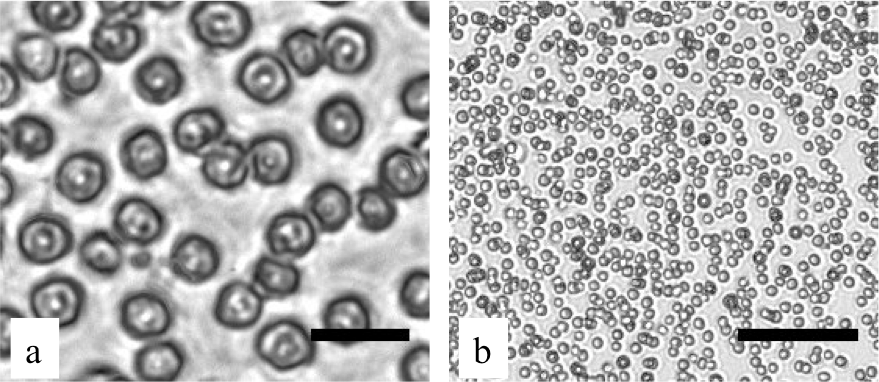}
\end{minipage}
\end{figure}

The RBC suspensions were prepared using horse blood (Fig.~\ref{fig:FigS1}). The blood sample was provided by TCS Biosciences (UK), in anticoagulant ethylenediaminetetraacetic acid (EDTA) 1.5 mg/mL, and stored at 4 $^\circ$C until further use. Sample processing and disposal were performed following the Ethics Committee for Animal Experimentation at University of Strathclyde. Since the blood was provided from different animals in distinct periods, the packed cell volume (PCV) or haematocrit ranges from 30 to 50\% according to the blood provider. This variability is overcome by producing cell suspensions from the cell sediment after washing and centrifugation. The haematocrit ($Ht$), defined by the cell volume fraction, is controlled to be $Ht \leq 1\%$. 

\begin{figure}[!b]
\begin{minipage}{0.68\linewidth}
\includegraphics[width=1.0\linewidth]{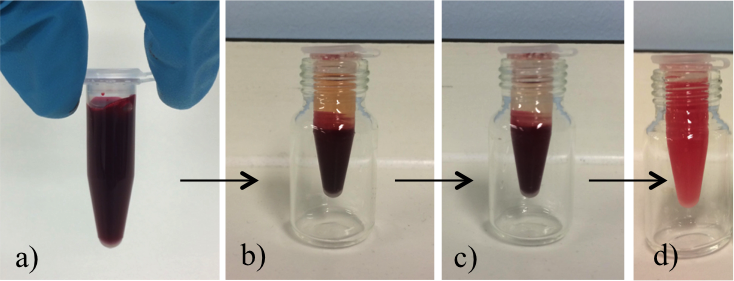}
\end{minipage}%%
\hfill
\begin{minipage}{0.30\linewidth}
\caption{\label{fig:FigS2} Steps involved in the preparation of RBC suspensions. (a) Eppendorf with whole blood; (b) separation (by centrifugation) of the supernatant plasma and buffy coat from the RBCs sediment; (c) transparent supernatant after RBC washing and centrifugation with PS; (d) RBC suspension in Dx40 solution ($Ht \leq 1\%$).}
\end{minipage}
\end{figure}

Fig.~\ref{fig:FigS2} presents the steps involved in sample processing. First, an Eppendorf with 1.5 mL of whole blood (WB) was centrifuged at 6000 rpm for 1 min (miniFUGE), after which the supernatant containing the yellow plasma and buffy coat (mainly white blood cells and platelets) was discarded. Second, the RBCs were washed twice with physiological saline PS (NaCl 0.9\% w/v), with the transparent supernatant discarded. Third, the final sample was prepared by suspending the desired volume of RBC in Dextran40 solution (Dx40 $M_w$ = 40000 g/mol, 0.1 g/mL in isotonic medium). This suspending medium will avoid the phenomenon of cell sedimentation during experiments, which is a common problem for suspensions prepared in PS or in original blood plasma.

\subsection{Data acquisition and image analysis}
\label{Data acquisition and image analysis, SI}

An inverted microscope (Olympus, IX71) was used to observe the device with bright field, employing a small magnification objective (10x/0.25 NA) in order to inspect cells across the entire channel depth and throughout a relatively large section of the channel length. \REV{We make use of a high precision syringe pump (neMESYS, Cetoni GmbH) of the low-pressure and 14:1 gear ratio type, which is designed to deliver accurate flow rates (on the order of nanoliter/min) for microfluidic applications. SGE gastight syringes of appropriate volume were used to ensure that the specified pulsation-free minimum dosing rate is always exceeded.} For each flow condition tested, the system was allowed to stabilize before image acquisition by a sensitive monochrome CCD camera (Olympus, XM10) at a frame rate of 15 Hz. The exposure time was set to 100 $\upmu$s allowing well-defined images of the cell boundaries. Videos of at least 300 frames each, corresponding to approximately 20 seconds of experiment, were acquired for four regions of interest (ROI). All the images were obtained at the channel centreplane ($z = 0$). All ROIs have a length of 130 $\upmu$m. The images acquired need post-processing for image enhancement and cell detection. For this purpose, we applied an automatic cell counting routine developed \textit{in-house} on ImageJ software.

\subsubsection{Background correction}
\begin{figure}[!h]
\begin{minipage}{0.5\linewidth}
\includegraphics[width=1.0\linewidth]{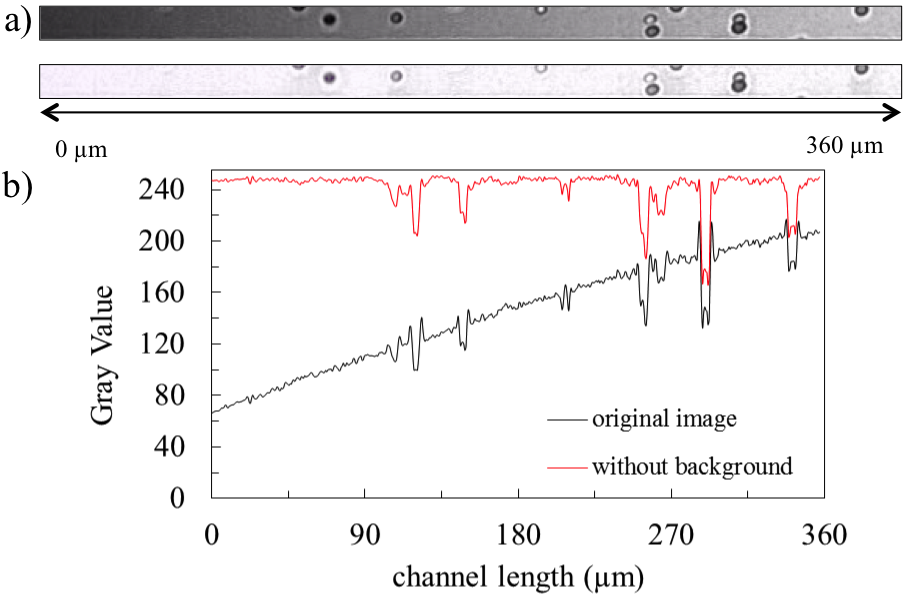}
\end{minipage}%%
\hfill
\begin{minipage}{0.45\linewidth}
\caption{\label{fig:FigS3} Background correction using Rolling Ball Radius function in ImageJ. (a) Image sections before and after background subtraction; (b) Corresponding intensity profiles to the image sections in (a).}
\end{minipage}
\end{figure}

The bright field images obtained from experiments usually present an unbalanced intensity distribution, in particular close to the channel inlets and outlets, where the fittings are connected to the device. To enhance the images, they were first corrected by subtracting the background. For this purpose, the Rolling ball radius filter was used, applying a radius of 10 pixels, which corresponds to an approximate diameter of a horse RBC ($D_{RBC} \approx 6\,\upmu$m). This procedure allowed obtaining a clear and balanced image without losing information of the RBC (Fig.~\ref{fig:FigS3}), especially for those cells located far away from the focal plane.

\subsubsection{Cell detection}

\begin{figure}[!t]
\centering
\includegraphics[width=1.0\linewidth]{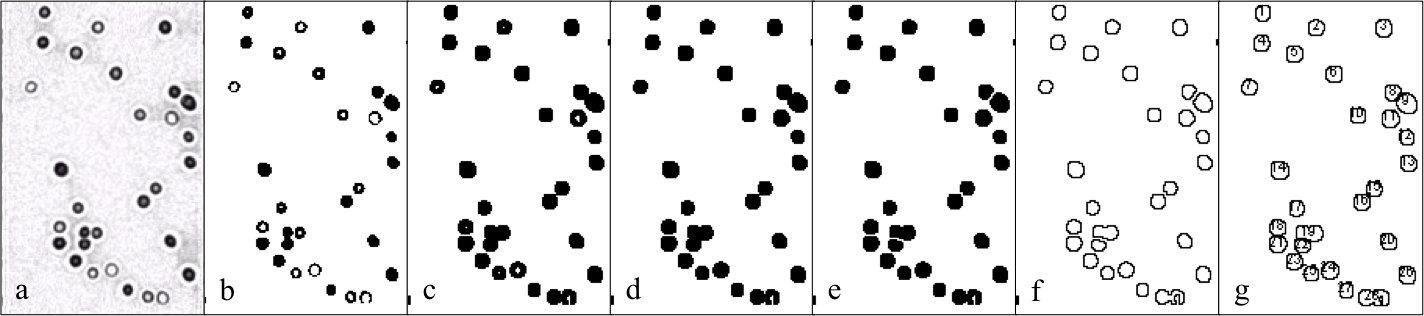}
\caption{\label{fig:FigS4} Procedures of detecting cells from the experimental images. (a) Image with corrected background; (b) The corresponding binary image; (c) function Dilate; (d) function Fill Holes; (e) function Watershed; (f) function Outline; (g) detected cells after function Analyse Particles on applying the Circularity filter.}
\end{figure}

The RBCs in images with corrected background are then detected and counted using an automatic routine developed on ImageJ open software. Fig.~\ref{fig:FigS4} describes the intermediate steps for further image enhancement and cell detection, based on the image with corrected background. First, a binary image (Fig.~\ref{fig:FigS4}b) is generated from the image with corrected background (Fig.~\ref{fig:FigS4}a), followed by function Dilate (Fig.~\ref{fig:FigS4}c) and function Fill holes (Fig.~\ref{fig:FigS4}d). This procedure allows recovering the cell contour for brighter cells that are located away from the focal plane. Subsequently, function Watershed is applied to generate separation among slightly overlapped cells (Fig.~\ref{fig:FigS4}e). Then, function Outline allows delimiting each foreground object in the binary image, by generating a one-pixel wide outline (Fig.~\ref{fig:FigS4}f). Based on this image, function Analyse Particles provides information of each delimited object, following which the Circularity filter is applied to detect objects whose circularity is in the range of 0.5$\sim$1.0, defined by $4\pi A/P^2$ (where A is the cell area and P the cell perimeter). The output is the image presented in Fig.~\ref{fig:FigS4}g, providing numbered cells together with a list of values including the center of mass ($X_{CoM}$, $Y_{CoM}$) and perimeter of each cell. A second filter is applied to the perimeter of the automatic detected object, in order to delete the ones presenting even smaller sizes than an undersized cell ($D_{RBC} \approx 4\,\upmu$m).

\subsection{Probability density distribution (PDD) of RBCs}
\label{Probability density distribution (PDD) of RBCs, SI}
\subsubsection{Calculation of local RBC fractions}
Local RBC fractions in experiments are determined using the center of mass information from detected cells within each ROI. We divide the whole channel width into a number of bins ($\geq$ 26) and count the number of cells that fall into each bin. The RBC fraction for each bin is then calculated as corresponding cell percentage (\%) relative to the total number of cells detected across the whole channel width. By plotting the RBC fractions on the cross-section of the designated ROI, the Probability density distribution (PDD) of RBCs is obtained, which can be regarded as an approximate representation of the haematocrit profile.

\subsubsection{Accuracy of the PDD profile}
To verify the accuracy of PDD profiles depicting RBC distribution obtained from the automatic counting method, two validation tests were performed. First, we examined the minimum number of frames needed to ensure a converged distribution. For a given ROI, the RBC distribution was determined using four different image stacks, containing 50, 100, 200 and 300 image frames, respectively. The PDD profiles resulting from a relatively small stack, e.g., 50 and 100 frames, were found to present significant noises and differ much from each other (Fig.~\ref{fig:FigS5}a). In contrast, the distributions obtained from a larger stack, namely 200 and 300 frames, were much smoother and appeared similar to each other (Fig.~\ref{fig:FigS5}a). Therefore, we conclude that a stack with 300 frames is enough to generate a converged RBC distribution. 

Second, we compared the RBC distribution obtained from our automatic counting using a stack of 300 frames with one based on manual counting. The result showed a high-level similarity between the two distributions (Fig.~\ref{fig:FigS5}b), hence validating the accuracy of our proposed method for the analysis of RBC distribution in experiments using low haematocrit ($Ht \leq 1\%$), where the boundaries of individual cells can be well defined. 

\subsubsection{Effect of cell overlapping}
\begin{figure}[!h]
\centering
\includegraphics[width=1.0\linewidth]{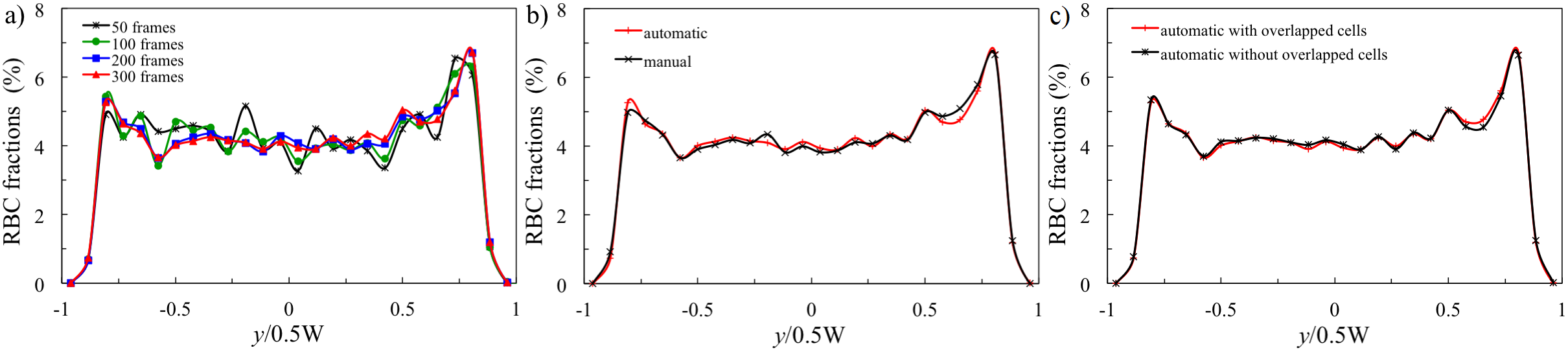}
\caption{\label{fig:FigS5} (a) Effect of the number of frames on the RBC distribution calculated by the automatic cell counting method. The PDD profiles are calculated using a stack of 50 (black line), 100 (green line), 200 (blue line) and 300 (red line) frames, respectively. (b) PDD profiles (using a stack of 300 images) obtained from automatic cell counting (red line) and manual cell counting (black line), respectively. (c) Effect of cell overlapping on the calculated RBC distribution, showing PDD profiles obtained from automatic counting with (red line) and without (black line) overlapped cells in the images, respectively.}
\end{figure}

The effect of cell overlapping is also investigated (Fig.~\ref{fig:FigS5}c). In Fig.~\ref{fig:FigS4}g, it is possible to observe false ``agglomerates'' (given the RBC concentration and the type of Dextran used in our experiments, we expect no physical aggregation) caused by the overlapping of cells across the channel depth direction. To study the impact of these ``agglomerates'' on the calculated RBC distribution, they were identified and excluded in a test case. The resulting RBC distribution was then compared to the original RBC distribution, where the two distributions obtained are found virtually identical (Fig.~\ref{fig:FigS5}c). This suggests that the statistical effect of cell overlapping is negligible on the final cell distribution obtained using our automatic counting method.

\REV{
\subsubsection{Effect of flow-inlet misalignment}
To test the effect of inlet conditions on the experimental results, the inlet ports are manually punched onto the PDMS, therefore the tubing connection can be either aligned or misaligned relative to the centreline of the microdevice. Here, Fig.~\ref{fig:FigS6} shows the evolution of PDD profiles under the misaligned configuration, corresponding to the channel in Fig.~\ref{fig:Fig1}c of the main text.

\begin{figure}[!h]
\begin{minipage}{0.55\linewidth}
\includegraphics[width=1.0\linewidth]{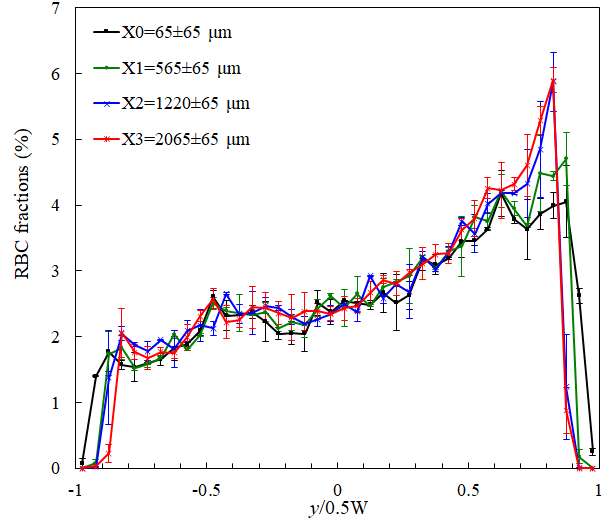}
\end{minipage}%%
\hfill
\begin{minipage}{0.38\linewidth}
\caption{\label{fig:FigS6} PDDs monitored at $x = 65\pm 65$, $565\pm 65$, $1220\pm 65$, $2065\pm 65~\upmu$m away from the entrance in the misaligned configuration, under a volume flow rate of $Q = 0.2\,\upmu$L/min.}
\end{minipage}
\end{figure}

\subsubsection{Effect of volume flow rate}
The volume flow rate was found to have an impact on the initial PDD profile at $x = 65\,\upmu$m. In general, a lower $Q$ (e.g., $Q = 0.2\,\upmu$L/min) yields a virtually flat density profile as  Fig.~\ref{fig:Fig2}a of the main text shows, whereas a higher $Q$ (e.g., $Q = 4.0\,\upmu$L/min) gives rise to appreciable density peaks near the walls as in Fig.~\ref{fig:FigS7}b here. This dependence of initial RBC distribution on $Q$ presumably arises because the cell deformation changes significantly within the investigated range of flow rates, thus non-trivially altering the lift forces imposed on the cell.

\begin{figure}[!h]
\begin{minipage}{0.75\linewidth}
\includegraphics[width=1.0\linewidth]{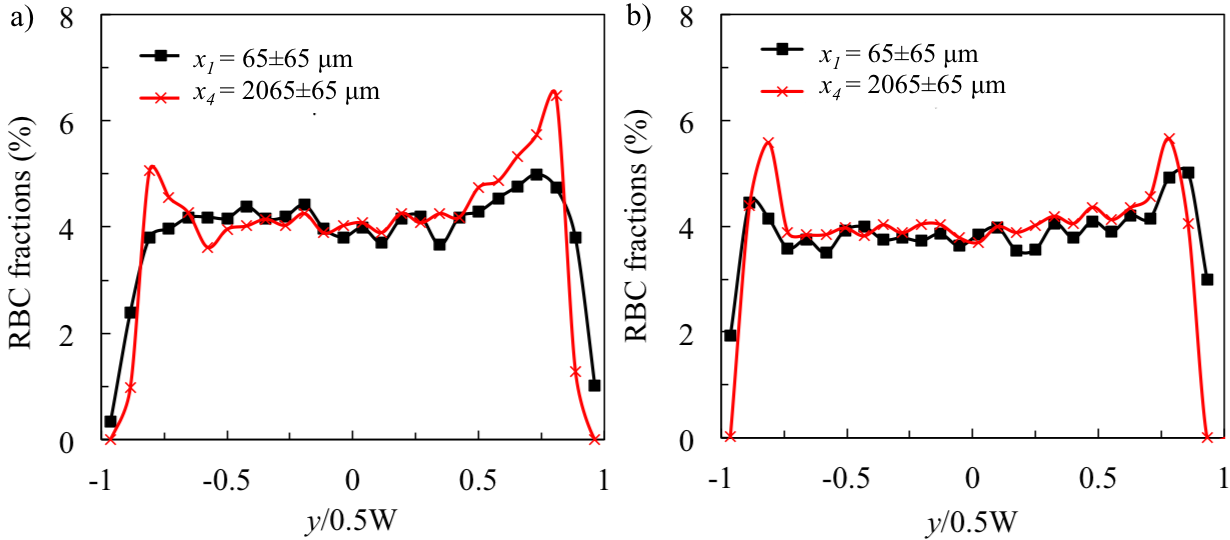}
\end{minipage}%%
\hfill
\begin{minipage}{0.24\linewidth}
\caption{\label{fig:FigS7} \REV{Experimental PDD profiles monitored at $x = 65$, $2065~\upmu$m for a $H_\text{F} =$ 1\% RBC suspension in the aligned configuration (see Fig.~\ref{fig:Fig1}b of the main text). The volume flow rates are (a) $Q = 0.8\,\upmu$L/min; (b) $Q = 4.0\,\upmu$L/min.}}
\end{minipage}
\end{figure}
}

\subsection{Cell-free layer (CFL)}
\label{Cell-free layer (CFL), SI}
\begin{figure}[!h]
\centering
\includegraphics[width=0.8\linewidth]{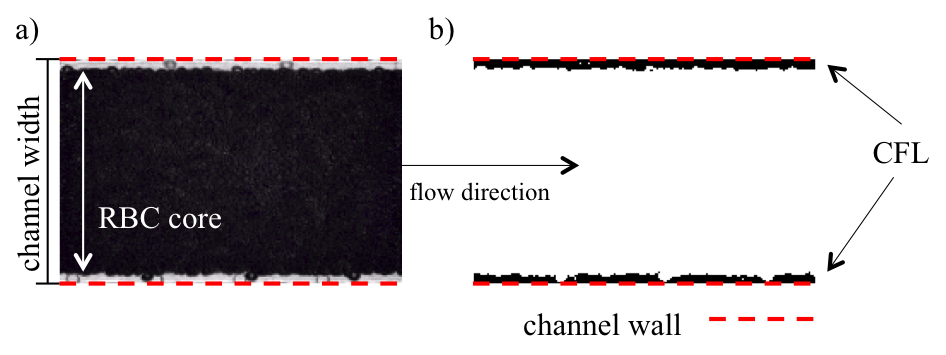}
\caption{\label{fig:FigS8} Determination of the CFL thickness. (a) Minimum intensity Z-projection image using a stack of 300 frames. (b) Binary image of the Z-projection for calculation of the CFL thickness.}
\end{figure}

To measure the CFL at a ROI within the channel, a combined image of 300 frames via minimum intensity Z-projection is generated (Fig.~\ref{fig:FigS8}a). Using this method, if a cell is passing through the ROI with lower intensity than the background, it will be registered; with a sufficient number of image frames, the layer depleted of cells (CFL) close to the walls can be distinguished. The binary image of such a projection (Fig.~\ref{fig:FigS8}b) presents a sharp boundary separating the RBC core and the CFL. The average CFL thickness is then defined as half of the averaged difference between the channel width and the RBC core along the ROI.

\begin{figure}[!h]
\centering
\includegraphics[width=0.8\linewidth]{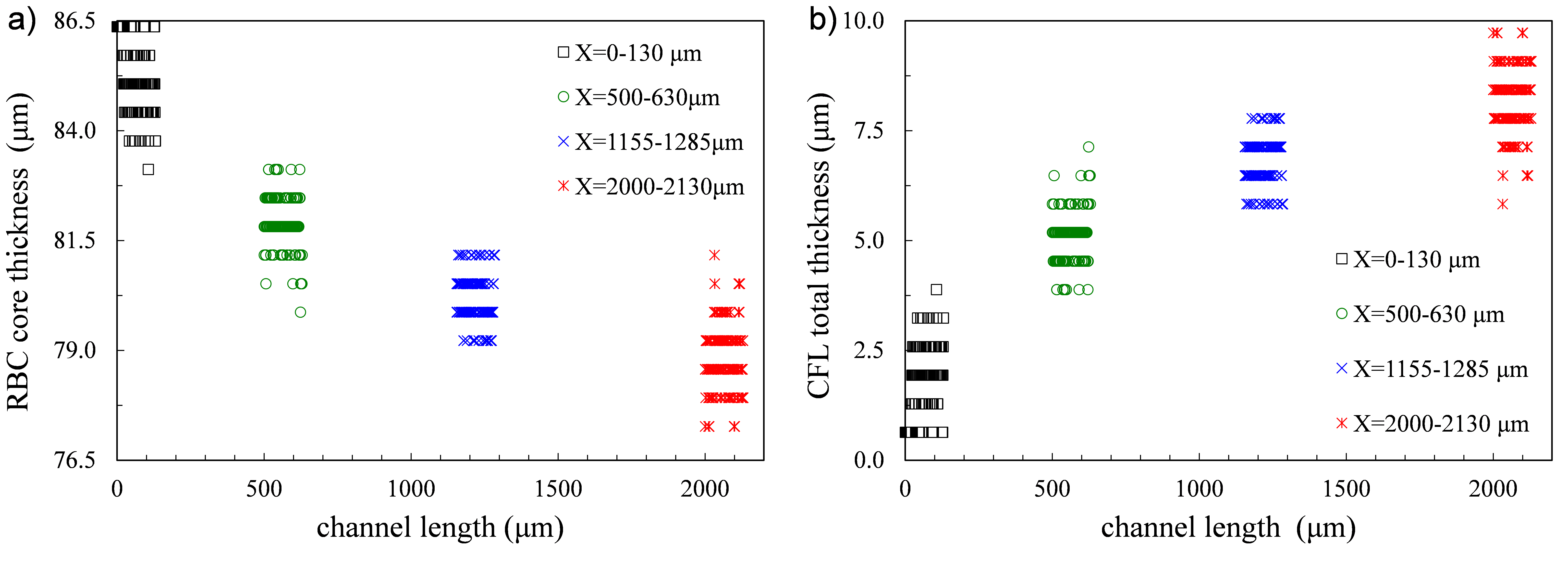}
\caption{\label{fig:FigS9} \REV{(a) Thickness of the RBC core measured at selected windows of the channel. (b) Calculated CFL values (accounting for both walls) based on RBC core thickness and channel width. (a--b) are for experiment $Q = 0.2\,\upmu$L/min.}}
\end{figure}

\section{Simulations}
\label{Simulations, SI}

\subsection{Model configuration}
\label{Model configuration, SI}

\begin{figure}[!t]
\centering
\includegraphics[width=0.75\linewidth]{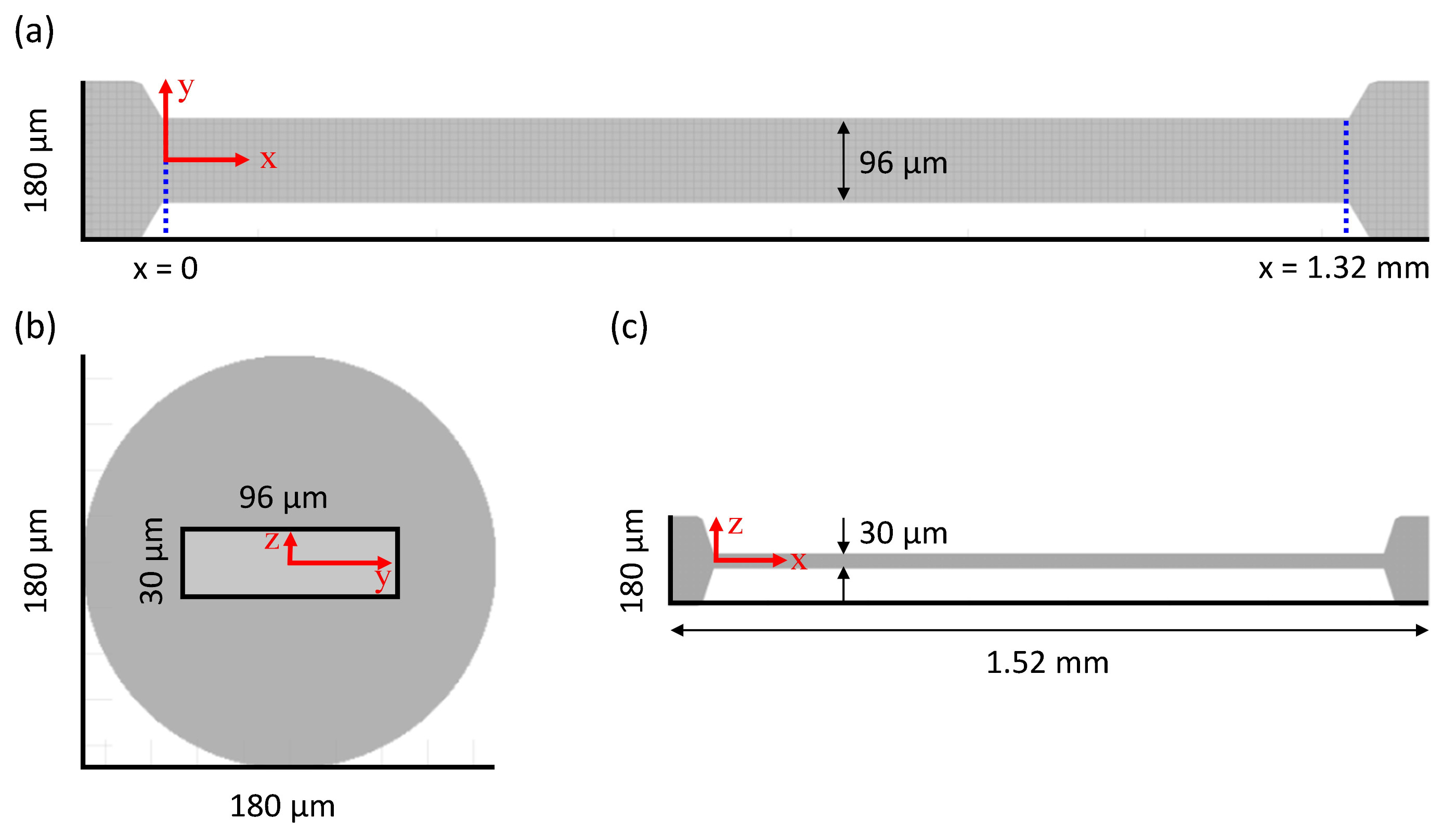}
\caption{\label{fig:FigS10} Geometry of the computational model (CTRAC-L, larger contraction). (a) is the top view of the geometry, showing the widthwise span of the channel. In (a), the main channel (rectangular) is marked with two blue dotted lines. (b) shows cross-sectional views of the front-end and rear-end of the tapered entry region, the latter of which coincides with the channel entrance ($x = 0$). (c) is the side view of the geometry, showing the depthwise span of the channel.}
\end{figure}

Fig.~\ref{fig:FigS10} shows the 3D computational model for simulating the RBC flow through a rectangular microchannel. Two geometries have been designed in the simulation to account for potential entrance effects that exist in experiments due to inlet configuration, which primarily arise from the inflow contraction between the entry region and the main channel. The main channel part in both geometries is the same, which is 96 $\upmu$m wide, 30 $\upmu$m deep and 1.32 mm long. The only difference is the front-end diameter of the tapered entry region (100 $\upmu$m long). The entry region for the first geometry has a front-end diameter of 96 $\upmu$m (hereafter referred to as CTRAC-S, meaning ``smaller (inflow) contraction''), and the one for the second geometry has a front-end diameter of 180 $\upmu$m (hereafter referred to as CTRAC-L, meaning ``larger (inflow) contraction''. All simulation data and plots presented in this paper are from CTRAC-L, unless otherwise specified). An identical entry region is also added to the channel outlet for removing cells from the simulation. The whole domain measures 1.52 mm in length, with a total of 26,201,764 lattice voxels (each $\Delta x$ = 2/3 $\upmu$m. All dimensions of the domain are labelled in Fig.~\ref{fig:FigS10}, shown from three camera views. 

\subsection{Simulation setup}
\label{Simulation setup, SI}

\begin{figure}[!t]
\centering
\includegraphics[width=0.75\linewidth]{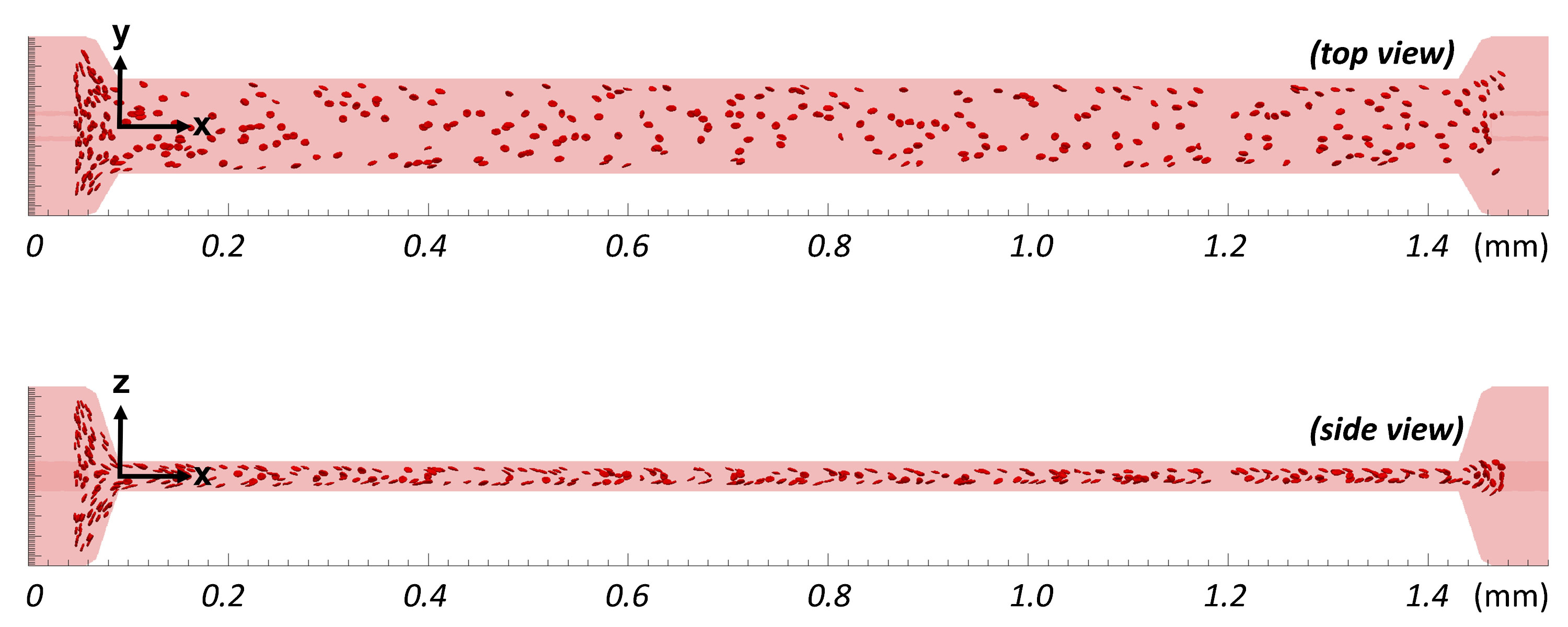}
\caption{\label{fig:FigS11} A snapshot of the simulation at t = 0.37s, showing distribution of the 1\% RBC suspension in the width and depth directions of the microchannel. Cells arriving at the end of the channel are removed from the system.}
\end{figure}

Initially, a plasma flow (no cells) is driven from left to right along the $x$-axis direction with a fixed volume flow rate (Fig.~\ref{fig:FigS11}). The flow is controlled by imposing a parabolic velocity profile (assuming Poiseuille flow) at the inlet and a reference pressure at the outlet. The no-slip condition on walls is implemented with the Bouzidi-Firdaouss-Lallemand method (2). Because it is computationally prohibitive to simulate the realistic experimental condition where the particle Reynolds number approaches zero, the fluid flow in our simulation is numerically accelerated to a state where the particle Reynolds number equals 0.03 (still within the low-inertia regime). After the plasma flow is converged, RBCs are randomly inserted from the cylindrical entry region in a continuous manner with fixed haematocrit ($H_\text{F}$ = 1\%). To make the experiments and simulations comparable, we establish the similitude law based on the capillary number $Ca$. For the target $Ca = 0.6$, we have fairly deformable RBCs in the channel assuming tank-treading motion, where the cell membrane is fluidized and rolls around the cell interior.

\subsection{Numerical analysis}
\label{Numerical analysis, SI}
\subsubsection{Calculation of RBC velocities}
\label{Calculation of RBC velocities, SI}
The lateral or axial velocity for an RBC is calculated as the unit displacement of its center of mass along the specific direction over a unit time step in the simulation:
\begin{equation} 
\label{eq:14}
V_{\text{l},y} = \Delta L_y/\Delta t, \quad V_{\text{l},z} = \Delta L_z/\Delta t, \quad V_\text{x} = \Delta L_x/\Delta t
\end{equation}
\noindent
Because the trajectories for all RBCs are recorded throughout the simulation, it is straightforward to group the cells into bins according to the location of their center of mass and conduct statistical analysis to estimate the time-average lift velocity (Fig.~\ref{fig:Fig4}) or streamwise velocity (Fig.~\ref{fig:FigS15}) of cells at a given position in the microchannel.

\begin{figure}[!t]
\centering
\includegraphics[width=0.75\linewidth]{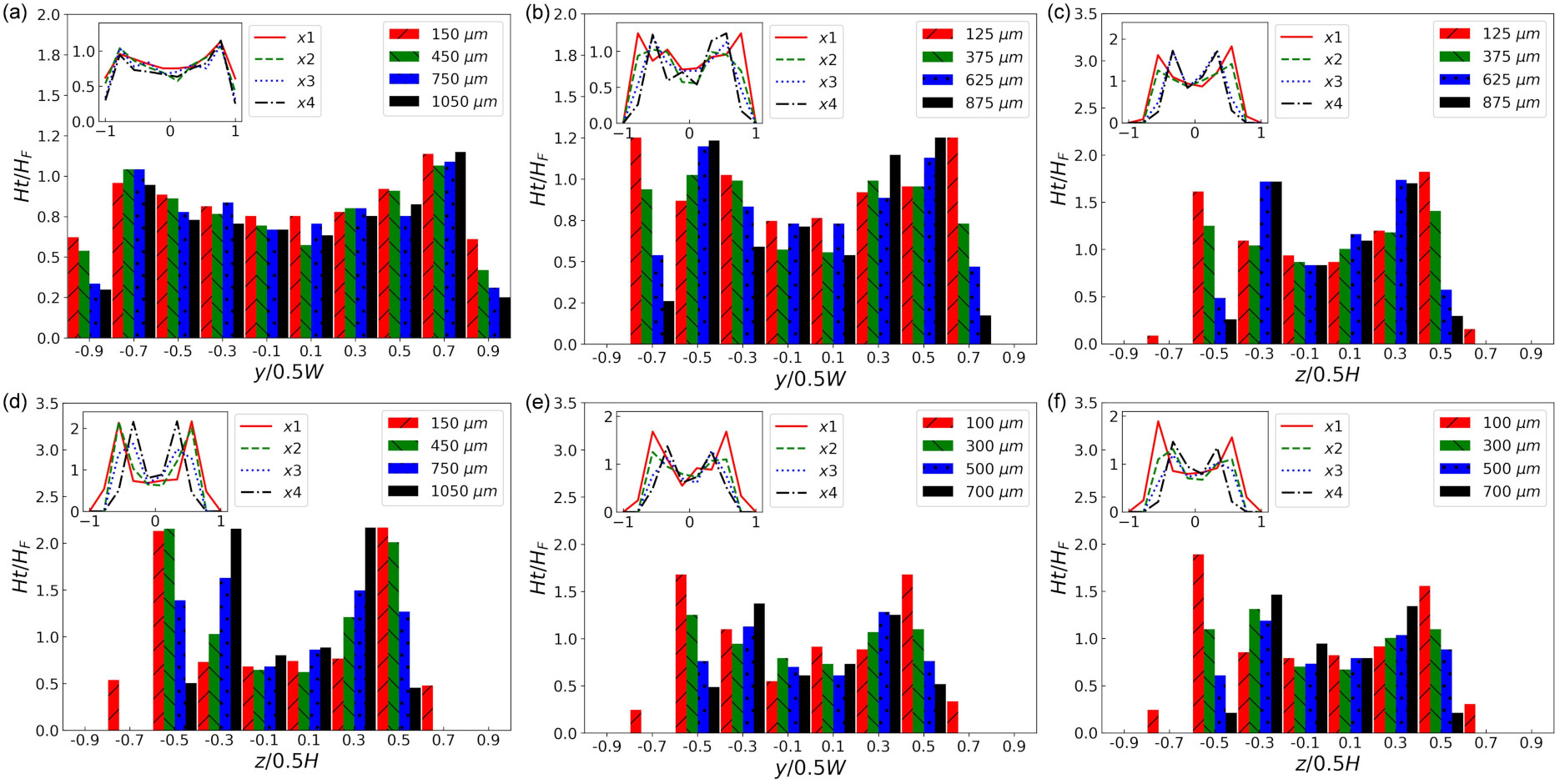}
\caption{\label{fig:FigS12} \REV{Haematocrit ($Ht$) histograms (bars) at $x = 3.3D_\text{h}$, $10D_\text{h}$, $16.5D_\text{h}$, $23D_\text{h}$ downstream of the channel, showing (a,b,c) the widthwise and (b,c,d) the depthwise distribution of RBCs at $H_\text{F} =$ 1\%. The inset in each panel shows the approximate $Ht$ profiles enveloping the calculated histograms for visual clarity. (a--b) are for a simulation in a channel with $AR = 3.2$ and $Ca = 0.1$. (c--d) are for a simulation in a channel with $AR = 1.7$ and $Ca = 0.6$. (e--f) are for a simulation in a channel with $AR = 1.0$ and and $Ca = 0.6$.}}
\end{figure}

\subsubsection{Haematocrit calculation} 
\label{Haematocrit calculation, SI}
Once the RBCs are fully perfused inside the channel and the number of cells within becomes stable, snapshots of the simulation are extracted at given time intervals, which are then assembled to count the cells for the calculation of time-averaged haematocrit profiles. For this dilute suspension, the number of time steps is set as $n_\text{t} = 100$ (with 2.22 ms interval between each) to ensure sufficient cells existing in the sampling pool. Next, various ROIs alongside the channel length are selected to count cells locally, starting at a position 150~$\upmu$m away from the entrance of the main channel. In total, 10 target positions between 150~$\upmu$m and 1050~$\upmu$m are chosen, with a 100~$\upmu$m increment. At each target position, we consider cells contained in a $W\times H\times\Delta L$ sampling box. All cells within this box are counted based on their centre of mass and allocated into a number of subdivisions either across the widthwise ($W$) direction or the depthwise ($H$) direction. The local haematocrits $Ht_{(i)}$ in each subdivision are then calculated as
\begin{linenomath}
\begin{equation} \label{eq:15}
Ht_{(i)} = \frac{n_{\text{rbc}(i)}Vol_\text{rbc}}{n_\text{t}Vol_\text{bin}}, \quad Vol_\text{bin} = \frac{Vol_\text{box}}{N_\text{bin}}, \quad (i=1,2,\ldots,N_\text{bin})
\end{equation}
\end{linenomath}
\noindent
where $n_{\text{rbc}(i)}$ is the number of cells detected in a given subdivision, $Vol_\text{rbc}=100$~fL is the standard volume of an RBC, $Vol_\text{bin}$ is the volume of each subdivision, $Vol_\text{box} = W\times H\times\Delta L$ ($\Delta L=50~\upmu$m) is the volume of the sampling box, and $N_\text{bin}=10$ is the number of subdivisions in $Vol_\text{box}$. \REV{Fig.~\ref{fig:FigS12} shows the obtained haematocrit histograms in both the width and depth directions of the channel from two simulations.}

\subsubsection{Calculation of RBC fluxes}
\label{Calculation of RBC fluxes, SI}
Based on the tube haematocrits $Ht_{(i)}$ (subdivided into $N_{bin}$ bins, Eq.~\ref{eq:15}) at a distance $x$ downstream of the flow inlet, the cross-sectional RBC fluxes $Q_{\text{rbc}(i)}$ allocated into the same number of bins can be calculated via the discharge haematocrit $H_{D(i)}$ and blood volume flow rate $Q_{\text{B}(i)}$:

\begin{equation} 
\label{eq:16}
Q_{\text{rbc}(i)} = H_{\text{D}(i)}Q_{\text{B}(i)}, \quad (i=1,2,...N_{\text{bin}})
\end{equation}

\begin{equation} 
\label{eq:17}
H_{\text{D}(i)} = Ht_{(i)}\left(\frac{\bar{V}_{\text{x}(i)}}{\bar{U}_{\text{B}(i)}}\right), \quad Q_{\text{B}(i)} = \bar{U}_{\text{B}(i)}A_{\text{bin}}
\end{equation}
\noindent
where $A_{\text{bin}}=W\times H/N_{\text{bin}}$ is the cross-sectional area of each bin, and $\bar{V}_{\text{x}(i)}$ and $\bar{U}_{\text{B}(i)}$ are the spatial-average RBC velocity and blood flow velocity within each bin, respectively. Combining Eqs.~(\ref{eq:16}) and (\ref{eq:17}), we have the following equation to be used for RBC flux calculation:
\begin{equation} 
\label{eq:18}
Q_{\text{rbc}(i)} = Ht_{(i)}\bar{V}_{\text{x}(i)}A_{\text{bin}}
\end{equation}

\subsubsection{Calculation of cell-free layer}
\label{Calculation of cell-free layer, SI}
To characterise the CFL in a rectangular channel with width $W$ and depth $H$, we consider a cross-section at a distance $x$ downstream of the flow inlet to be divided into four triangles by the diagonals and the wall edges. The heights of the four triangles, each perpendicular to a channel wall, are located at $\theta = 0, \pi/2, \pi, 3\pi/2$, respectively. We define the RBC density $\phi(h,\theta',x,t)$ at a point $p(x,h,\theta')$ of longitudinal coordinate $x$, height coordinate $h$ and angular coordinate $\theta'$ within a given triangle for a certain time instant $t$ as 1 if that point is enclosed by at least one RBC membrane; otherwise, $\phi(h,\theta',x,t) = 0$. The time-mean RBC density within any of the fours triangles can then be obtained via integration over $(h, \theta')$ and average over $t$:
\begin{equation} 
\label{eq:19}
\bar\phi(x,\theta) = \frac{1}{N}\sum_{i=1}^{N}\int_{0}^{L_\text{h}}\int_{\theta-\Theta/2}^{\theta+\Theta/2} \phi(h,\theta',x,t_i)h\tan|\theta'-\theta| \text{d}\theta' \text{d}h
\end{equation}
\noindent
wherein $N$ is the number of time steps extracted from the simulation, $L_\text{h}$ is the height of the triangle and $\Theta$ is its tip angle centred in the cross-section. For the left and right triangles:
\begin{equation} 
\label{eq:20}
L_\text{h} = W/2, \quad \Theta = 2\arctan(H/W)
\end{equation}
\noindent
and for the top and bottom triangles:
\begin{equation} 
\label{eq:21}
L_\text{h} = H/2, \quad \Theta = 2\arctan(W/H)
\end{equation}

\begin{figure}[!h]
\centering
\includegraphics[width=0.95\linewidth]{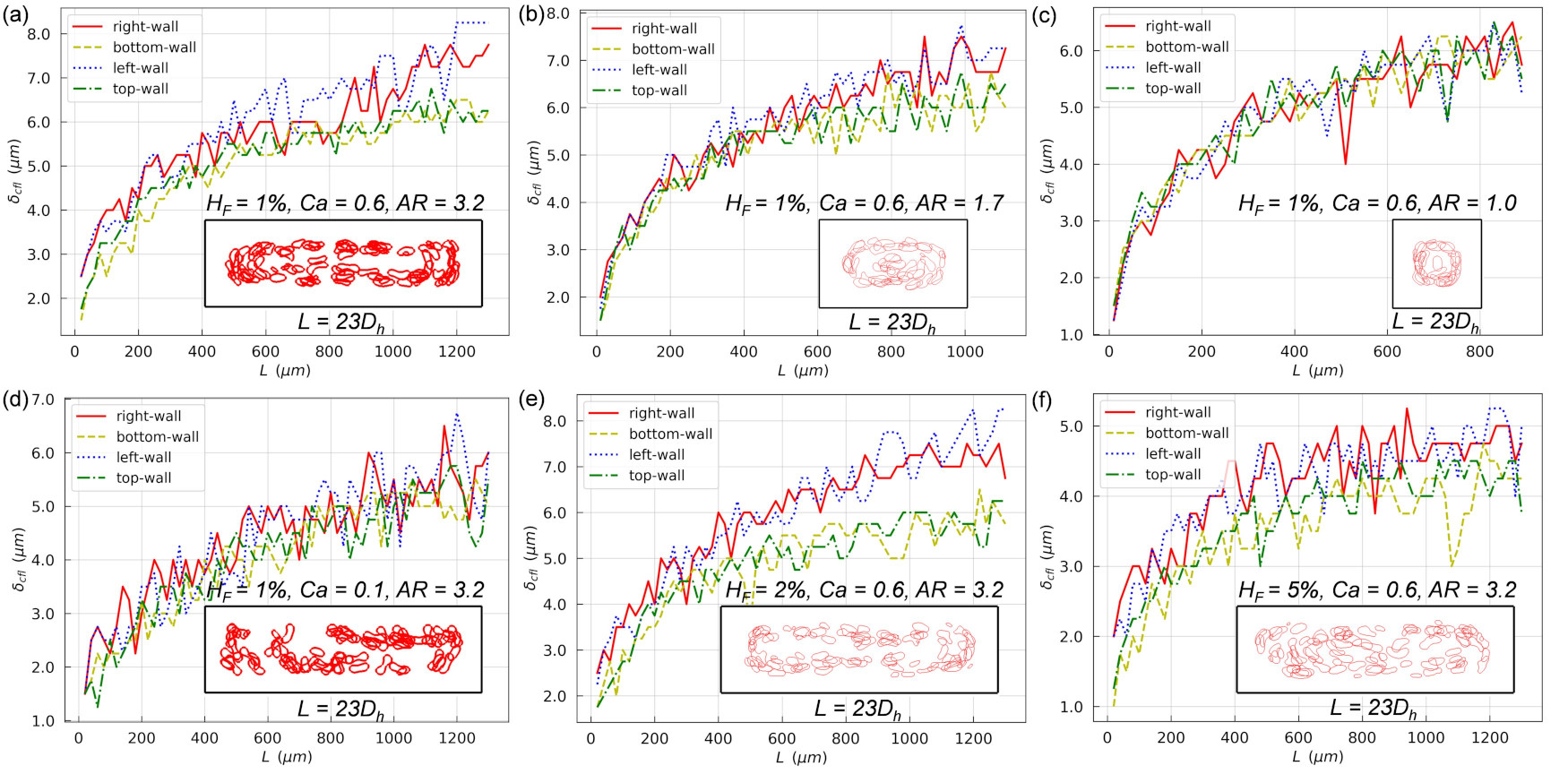}
\caption{\label{fig:FigS13} \REV{Numerical CFLs calculated along each wall of the channel (Right/Bottom/Left/Top-walls) from different simulations. (a) $H_\text{F} =$ 1\%, $Ca = 0.6$, AR = 3.2; (b) $H_\text{F} =$ 1\%, $Ca = 0.6$, AR = 1.7; (c) $H_\text{F} =$ 1\%, $Ca = 0.6$, AR = 1.0; (d) $H_\text{F} =$ 1\%, $Ca = 0.1$, AR = 3.2; (e) $H_\text{F} =$ 2\%, $Ca = 0.6$, AR = 3.2; (f) $H_\text{F} =$ 5\%, $Ca = 0.6$, AR = 3.2. The inset in each panel shows a cross-sectional view of the RBC pattern at $L = 23D_\text{h}$ away from the entrance, combining snapshots from 50 consecutive time steps from the corresponding simulation.}}
\end{figure}

\begin{figure}[!h]
\centering
\includegraphics[width=0.95\linewidth]{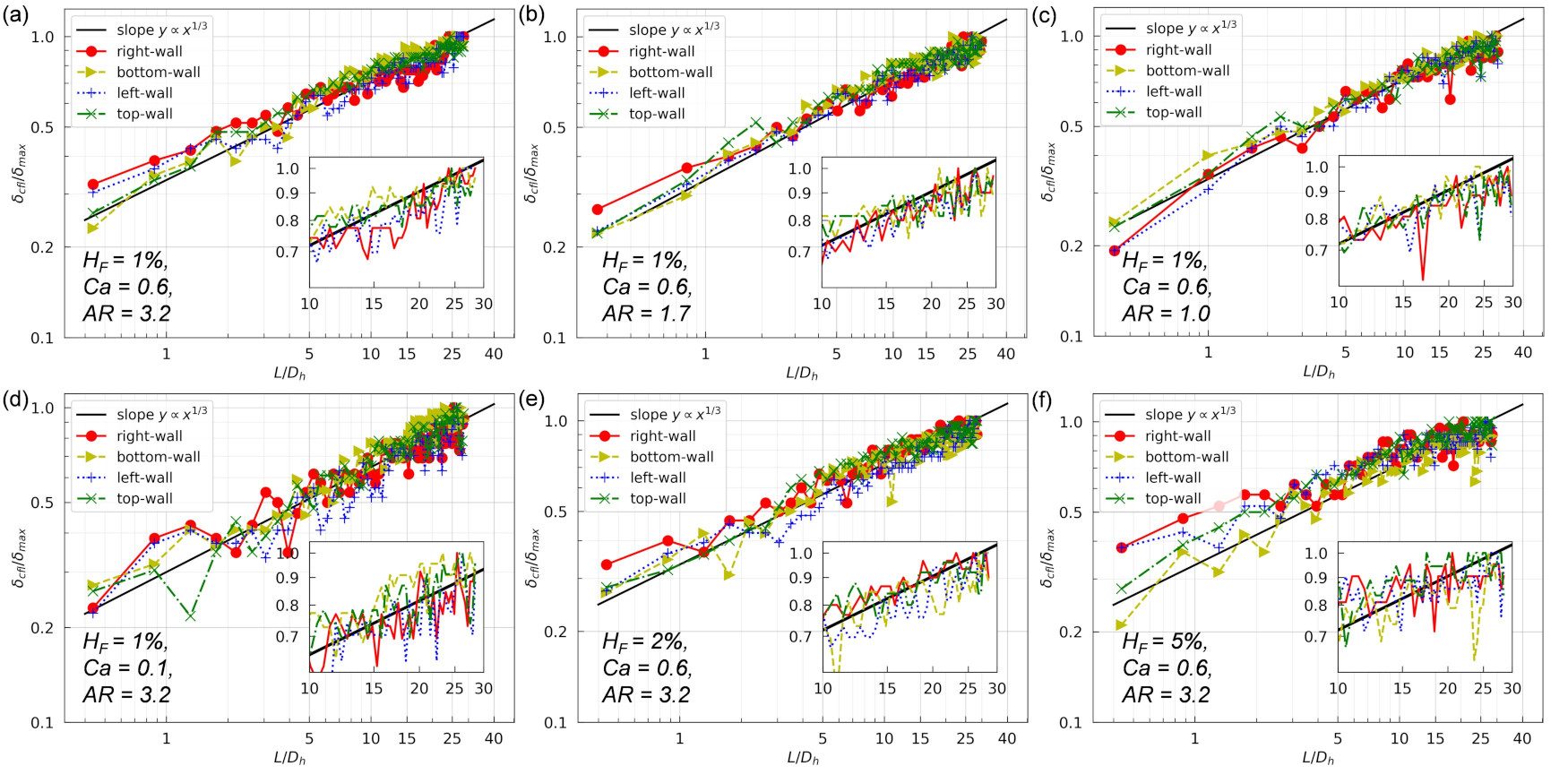}
\caption{\label{fig:FigS14} \REV{Log-log plots of the CFLs as in Fig.~\ref{fig:FigS13} against a 1/3 power-law trendline. The CFL values $\delta_\text{cfl}$ are normalised by the maximum CFL thickness $\delta_\text{max}$ detected within the whole investigated range, and the development lengths $L$ are normalised by the channel hydraulic diameter $D_\text{h}$. The inset in each panel show a zoom-in view of the range $L = 10\sim28 D_h$.}}
\end{figure}

If we define the CFL as a thin layer in proximity to a given wall (base of the triangle) which contains only a negligible fraction of RBC density $\varepsilon$, e.g., $\varepsilon < 1e-6$, then the thickness of this layer $\delta(x,\theta)$ can be determined using numerical iterations:
\begin{equation} 
\label{eq:22}
\frac{1}{\bar\phi(x,\theta)N}\sum_{i=1}^{N}\int_{L_\text{h}-\delta(x,\theta)}^{L_\text{h}}\int_{\theta-\Theta/2}^{\theta+\Theta/2} \phi(h,\theta',x,t_i)h\tan|\theta'-\theta| \text{d}\theta' \text{d}h \geq \varepsilon
\end{equation}
\noindent
By calculating the threshold thickness $\delta(x,\theta)$ satisfying the above criteria for each $\theta$ at consecutive $x$ locations, \REV{we can obtain the CFL values for each wall along the channel-axis direction as in Fig.~\ref{fig:FigS13}. The development of these CFLs is further plotted in log-log scale against a 1/3 slope indicating power-law growth (Fig.~\ref{fig:FigS14}).}

\subsection{Geometric effect and shear anisotropy}
\label{Geometric effect and shear anisotropy, SI}

\begin{figure*}[!t]
\centering
\includegraphics[width=1.0\linewidth]{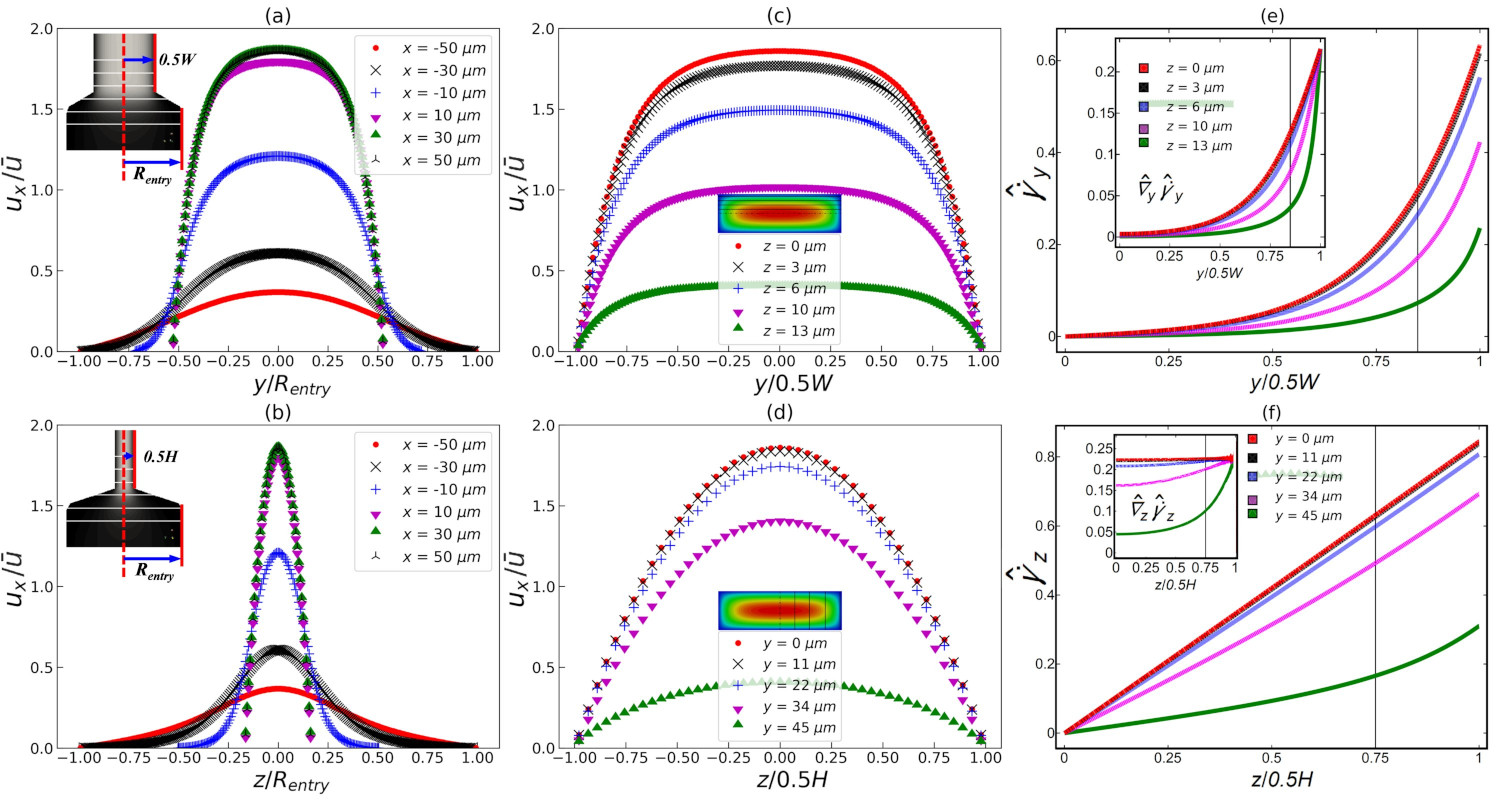}
\caption{\label{fig:FigS15} 3D flow in the rectangular channel. (a--b) (Simulation) Transient mid-plane velocity profiles near the entrance at positions labelled by white solid lines in the channel sketch. (c--d) (Simulation) Steady velocity profiles at $x = 50\,\upmu$m. $z$ and $y$ values in the legends represent positions relative to the mid-plane in each direction, with $z = 0$ and $y = 0$ denoting the widthwise and depthwise mid-planes, respectively. (e--f) (Analytical) Shear rates (main frame) and shear-rate gradients (inset) calculated from asymptotic solutions in Ref. (3). The colors represent $z$ or $y$ values as in (c--d). The black vertical lines in (e--f) indicate the center of the outmost layer of cells at $x = 50\,\upmu$m detected in the RBC simulation along the (e) $y$-direction and (f) $z$-direction, respectively (see Fig.~\ref{fig:Fig4}).}
\end{figure*}

In our problem, where a laminar flow is pressure-driven in a rectangular microchannel of high aspect ratio ($W/H=3.2$), the flow field is complex because distinct velocity profiles co-exist in the channel cross-section: a typical parabola characteristic of Poiseuille flow in the depth ($H$) direction and a blunted profile featuring plug-like flow in the width ($W$) direction (3). Additionally, the existence of entrance effects further increases the complexity of the problem. These two aspects contribute to a plethora of scenarios for the velocity profile curvature (reflecting the shear-rate gradient), which on the one hand constitutes a mechanism for cell migration itself (by introducing symmetry-breaking deformation, see Ref. (4)) and on the other hand impacts the wall-induced lift (being a function of the particle shape, see Ref. (5)). We show the evolution of unperturbed mid-plane velocity profiles (in the absence of cells) near the entrance across both directions, where rich behaviour of the velocity profile can be found as the fluid accelerates through the contracted transition between the entry region and the channel itself (Fig.~\ref{fig:FigS15}a--b). This geometry-induced velocity alteration potentially underpins the distinct initial distribution of RBCs between the $W$ and $H$ directions of the channel observed in Fig.~\ref{fig:Fig2}a,d.

Fig.~\ref{fig:FigS15}a--b reveal that the flow becomes fully-developed between $x$ = 30$\sim$50$~\upmu$m in our simulations. Fig.~\ref{fig:FigS15}c--d show the converged velocity profiles velocity profiles for different z- and y-planes at $x = 50\,\upmu$m, respectively. Using the asymptotic solution of this microfluidic flow (3), we further calculate and plot the shear-rate profiles and shear-rate-gradient profiles. The cross-sectional pattern of shear rates (normalised by the time scale $\tau$ as in Eq.~\ref{eq:2}) turns out more intricate than often assumed (Fig.~\ref{fig:FigS15}e--f), even in the narrow dimension ($H$) where linearity is commonly assumed for such a high aspect ratio ($W/H > 3$) (6). At close distances to the wall (e.g., $y=45\,\upmu$m in Fig.~\ref{fig:FigS15}d), the $H$-direction velocity profile deviates significantly from a parabolic one (Fig.~\ref{fig:FigS15}f), resulting in non-linear decrease of shear rate on approaching the channel centreline and varying shear gradients (inset of Fig.~\ref{fig:FigS15}f). In the $W$ direction where blunted velocity profiles prevail (Fig.~\ref{fig:FigS15}c), the shear gradients vary both at close and large distances to the wall (inset of Fig.~\ref{fig:FigS15}e), leaving the hydrodynamic lift even more unpredictable by the classic theory derived on the basis of ideal shear flows (4).

\subsection{Disturbance of RBCs to the plasma flow}
\label{Disturbance of RBCs to the plasma flow, SI}

\begin{figure*}[!t]
\begin{minipage}{0.68\linewidth}
\includegraphics[width=1.0\linewidth]{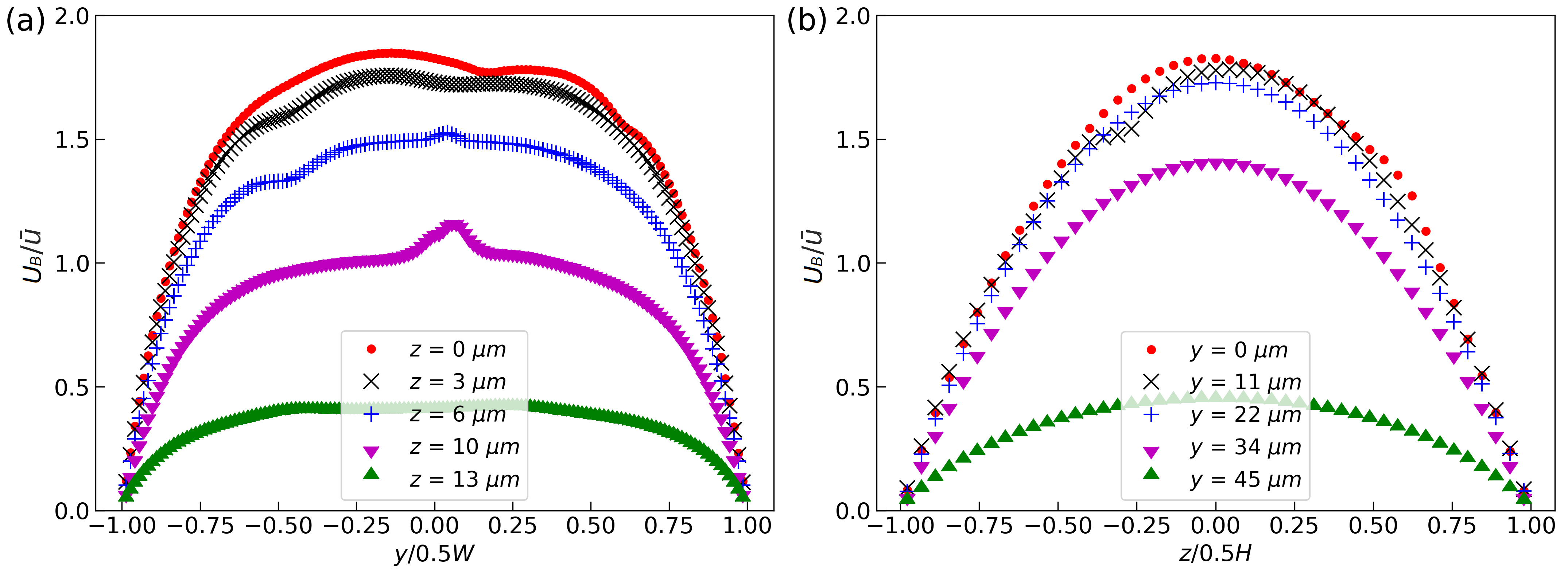}
\end{minipage}%%
\hfill
\begin{minipage}{0.30\linewidth}
\caption{\label{fig:FigS16} Velocity profiles of the RBC-laden flow at $x = 50\,\upmu$m. $z$ and $y$ values in the legends represent positions relative to the mid-plane in each direction, with $z = 0$ and $y = 0$ denoting the widthwise and depthwise mid-planes, respectively. The blood flow velocity $U_\text{B}$ is normalised by mean velocity of the unperturbed plasma flow $\bar{u}$.}
\end{minipage}
\end{figure*}

The velocity profiles from the RBC-laden flow at $Ht$ 1\% (Fig.~\ref{fig:FigS16}) apparently do not deviate much from the unperturbed flow in the absence of RBCs (Fig.~\ref{fig:FigS15}c,d). This implies that the sparse distribution of RBCs in the dilute limit only affects the flow velocity to a limited amount and causes local disturbances without modifying the overall velocity profile significantly. Therefore, it is reasonable to analyse the migration of RBCs in the 1\% suspension using the unperturbed plasma flow.

\subsection{RBC lateral migration and axial acceleration}
\label{RBC lateral migration and axial acceleration, SI}

\subsubsection{Cross-sectional RBC flux distribution}
\begin{figure*}[!h]
\centering
\includegraphics[width=0.68\linewidth]{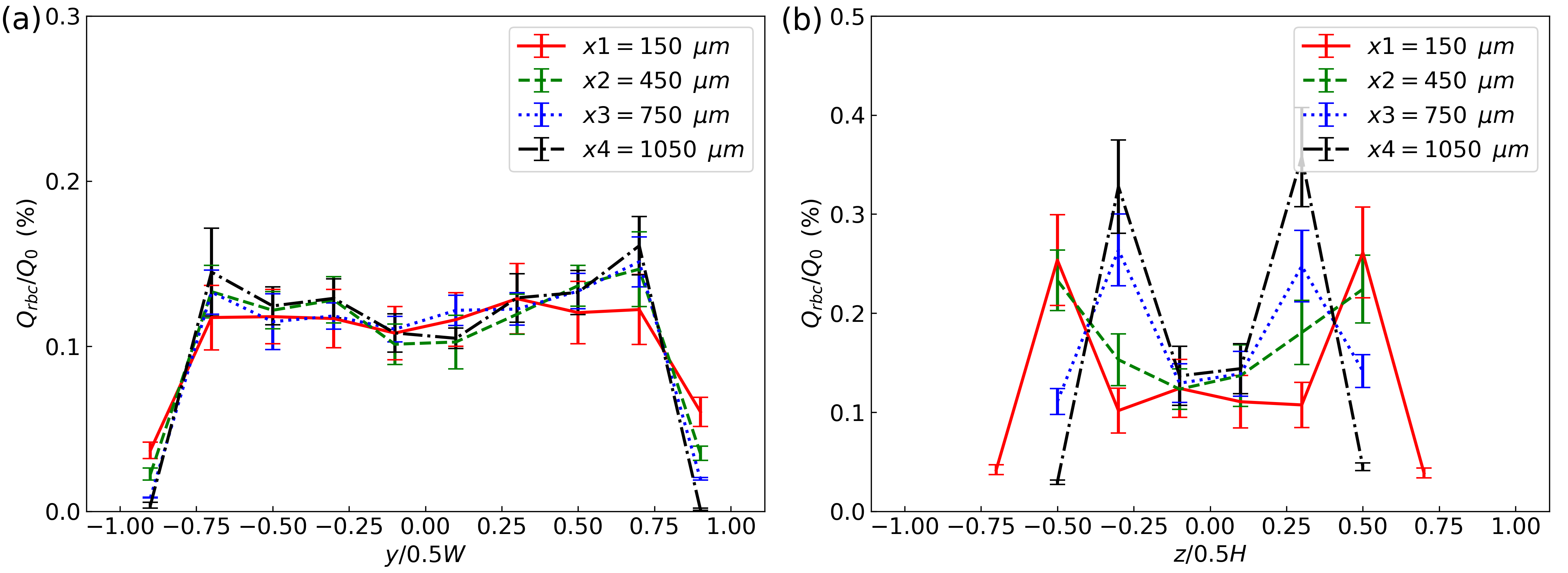}
\caption{\label{fig:FigS17} (Simulation) Evolution of RBC fluxes $Q_\text{rbc}$ across the channel in (a) the width direction and (b) the depth direction. The four colors red, green, blue, black indicate a distance of $x = 150, 450, 750, 1050\,\upmu$m away from the entrance, respectively. All RBC fluxes are normalised by the volume flow rate of the unperturbed flow $Q_0$ and expressed in percentages.}
\end{figure*}

The $Ht$ profile is co-determined by the lateral migration (toward channel centreline) and the streamwise acceleration (along channel axis) of RBCs, the former of which arises from the hydrodynamic lift effect and the latter is associated with the F\aa hr\ae us effect. Both effects have an impact the distribution of local haematocrits and can affect the dynamical evolution of the $Ht$ profile. In this sense, the peaks emerging in the $Ht$ profile (Fig.~\ref{fig:Fig2}) may not be ideal markers of the ``pseudo-equilibrium'' position for the inward migration of cells. Therefore, to ascertain the ``pseudo-equilibrium'' positions suggested by $Ht$ peaks, these two effects need to be decoupled. To this end, we calculate the RBC fluxes, the sum of which strictly conserves along the channel axis and examine the cross-sectional distribution of the RBC fluxes. This allows us to exclude the effect of cell acceleration and distinguish the amount of lateral migration of the cells (Fig.~\ref{fig:FigS17}).

The evolution of RBC-flux distribution turns out qualitatively similar to that of the $Ht$ profiles. It reveals a clear trend of inward RBC migration over time, with near-wall RBC fluxes moving away and subsequently accumulating at a ``pseudo-equilibrium'' position in both directions of the channel. For the widthwise migration, this equilibrium position is found to be 0.3 times of half channel width away from the wall; for the depthwise migration, it is 0.7 times of half channel away. The inward shift of RBC fluxes is accompanied by the continuous expansion of a zero-flux zone in the vicinity of the wall, which corresponds to the gradually growing CFL.

\subsubsection{Axial acceleration of RBCs}
\begin{figure*}[!h]
\centering
\includegraphics[width=0.68\linewidth]{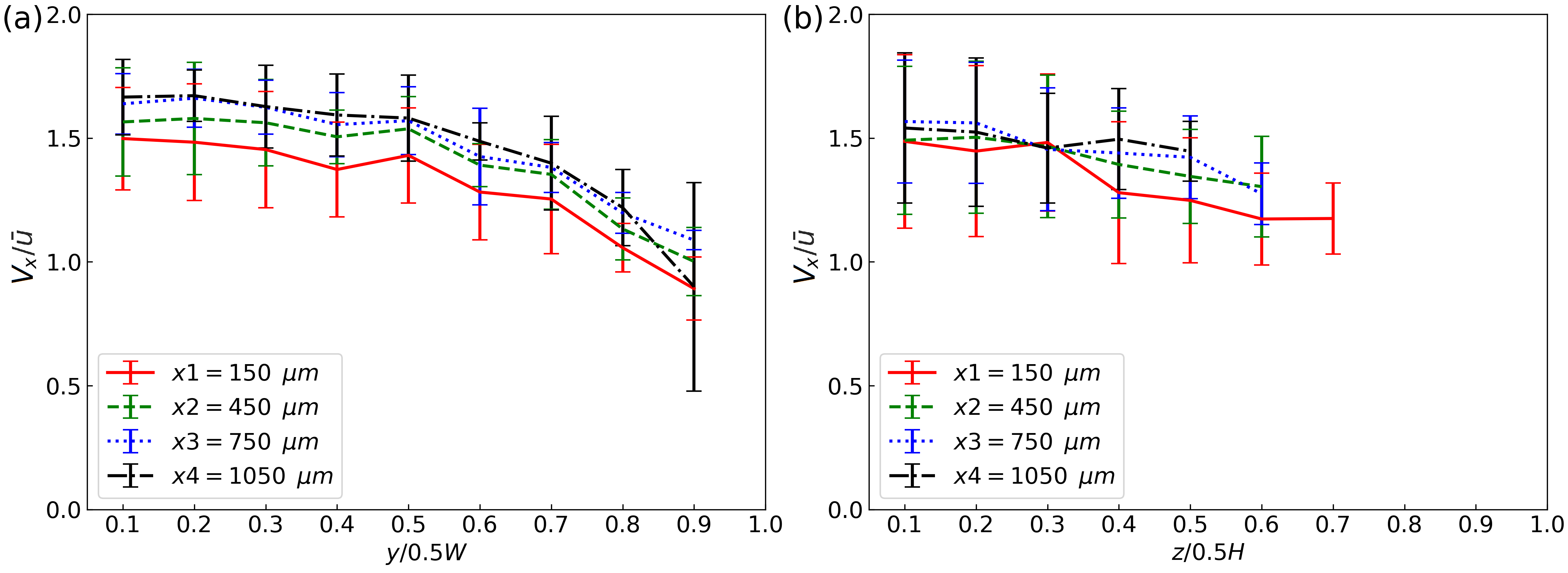}
\caption{\label{fig:FigS18} (Simulation) Axial velocity $V_x$ (streamwise, normalised by mean velocity of the unperturbed flow $\bar{u}$) of RBCs in the channel at $x = 150, 450, 750, 1050\,\upmu$m away from the entrance. (a) and (b) show the width and depth directions of the channel, respectively. For statistical analysis in both (a) and (b), the channel is fictitiously folded along the channel centreline and then divided into 10 bins, followed by the assignment of cells according to the position of their center of mass. Each error bar here shows the mean and standard deviation of the streamwise velocity.}
\end{figure*} 

\section*{Supporting references}
\label{SR}
\noindent 1. Losserand, S., G. Coupier, and T. Podgorski, 2019. Migration velocity of red blood cells in microchannels. \textit{Microvascular Research} 124:30-36.

\noindent 2. Bouzidi, M., M. Firdaouss, and P. Lallemand, 2001. Momentum transfer of a Boltzmann-lattice fluid with boundaries. \textit{Physics of Fluids} 13:3452–3459.

\noindent 3. Bruus, H., 2008. Theoretical Microfluidics. Oxford Master Series in Physics. OUP Oxford.

\noindent 4. Kaoui, B., G. H. Ristow, I. Cantat, C. Misbah, and W. Zimmermann, 2008. Lateral migration of a two-dimensional vesicle in unbounded Poiseuille flow. \textit{Physical Review E} 77:021903.

\noindent 5. Olla, P., 1997. The lift on a tank-treading ellipsoidal cell in a bounded shear flow. \textit{Journal de Physique II} 7:1533–1540.

\noindent 6. Coupier, G., B. Kaoui, T. Podgorski, and C. Misbah, 2008. Noninertial lateral migration of vesicles in bounded Poiseuille flow. \textit{Physics of Fluids} 20:111702.

\end{document}